\documentclass[a4paper,11pt]{article}
\usepackage{jheppub} 
 \usepackage{amssymb,amsthm}
 \usepackage{amsmath,amssymb,amsfonts,bm,amscd}
 \usepackage{graphicx}
 
 \usepackage{array}
 
 \usepackage{caption, subcaption}
 
 \usepackage{tikz}

 \usepackage{enumitem}
 \usepackage{tcolorbox}
 \usepackage{multirow}
 
\def\e{\epsilon}

\def\h{\theta}

\def\CA{{\cal A}}
\def\CB{{\cal B}}
\def\CC{{\cal C}}

\def\CE{{\cal E}}

\def\CI{{\cal I}}

\def\CM{{\cal M}}
\def\CN{{\cal N}}
\def\CO{{\cal O}}

\DeclareMathOperator{\Tr}{Tr}

\def\beq#1\eeq{\begin{align}#1\end{align}}

\usepackage{enumitem}

\def\fg{\mathfrak{g}}

\def\bfz{\mathbf{z}}

\def\half{\frac{1}{2}}

\newcommand{\pe}[1]{\mathrm{PE}\left[#1\right]}

\newcommand{\BB}{{\cal B}}
\newcommand{\DD}{{\cal D}}
\newcommand{\EE}{{\cal E}}
\newcommand{\QQ}{{\cal Q}}
\newcommand{\II}{{\cal I}}
\newcommand{\suup}{1}
\newcommand{\sudown}{2}

\global \long \def \NN{ \mathcal{N}}
\global \long \def \II{ \mathcal{I}}

\def\e{\epsilon}

\def\h{\eta}

\def\CT{{\cal T}}

\def\CI{{\cal I}}

\def\CN{{\cal N}}

\def\half{{\frac12}}

\def\IC{\relax\hbox{$\inbar\kern-.3em{\rm C}$}}

\def\IC{{\bf C}}

\def\CN{{\cal N}}

\def\bea{\begin{eqnarray}}
\def\eea{\end{eqnarray}}
\def\be{\begin{equation}}
\def\ee{\end{equation}}
\def\ba{\begin{align}}
\def\ea{\end{align}}
\def\bse{\begin{subequations}}
\def\ese{\end{subequations}}
\def\1F1{{}_1\!F_1}
\def\2F0{{}_2\!F_0}

\def\h3{$\textrm{H}_3^+$}

\def\IC{{\mathbb C}}

\def\Tr{{\rm Tr}}

\def\lbldef#1#2{\expandafter\gdef\csname #1\endcsname {#2}}

\def\href#1#2{#2}

\newcommand{\eeq}{\end{equation}}
\newcommand{\ber}{\begin{eqnarray}}
\newcommand{\eer}{\end{eqnarray}}

\def\be{\begin{eqnarray}}
\def\ee{\end{eqnarray}}

\providecommand{\tabularnewline}{\\}

\def\({\left(}
\def\){\right)}
\def\[{\left[}
\def\]{\right]}
\def\<{\langle}
\def\>{\rangle}

\title{A study of $\mathcal{N}=1$ SCFT derived from $\mathcal{N}=2$ SCFT: index and chiral ring}

\author[a,b]{Dan Xie}
\author[a]{ Wenbin Yan}

\affiliation[a]{Yau Mathematics Science center, Tsinghua University, Beijing, 100084, China}
\affiliation[b]{Department of Mathematics, Tsinghua University, Beijing, 100084, China}

\abstract{One can derive a large class of new $\mathcal{N}=1$ SCFTs by turning on $\mathcal{N}=1$ preserving deformations for $\mathcal{N}=2$ Argyres-Dougals theories. In this work, we use   $\mathcal{N}=2$ superconformal indices to get   indices of $\mathcal{N}=1$ SCFTs,  then use these indices to derive  chiral rings of $\mathcal{N}=1$ SCFTs.  For a large class of $\mathcal{N}=2$ theories, we find that the IR theory contains only free chirals if we deform the parent $\CN=2$ theory using the Coulomb branch operator with smallest scaling dimension. Our results provide interesting lessons on studies of $\mathcal{N}=1$ theories, such as $a$-maximization, accidental symmetries, chiral ring, etc.}

\begin{document} 
\maketitle
\flushbottom

\section{Introduction}
\label{sec:intro}
The study of 4d $\mathcal{N}=1$ supersymmetric field theory is quite important for understanding general quantum field theory. On one hand these theories have  richer
dynamics than theories with more supersymmetries \cite{Intriligator:1995au}, yet on the other hand these theories have minimal supersymmetry and so one can still get interesting  exact results.  We are particularly interested in 4d 
$\mathcal{N}=1$ superconformal field theory (SCFT). These theories can be realized as the IR fixed points of asymptotically free gauge theories with $\mathcal{N}=1$ supersymmetry \cite{Seiberg:1994pq}, or
obtained by using geometric method such as brane construction \cite{Giveon:1998sr,Witten:1997ep}, 6d $(2,0)$ construction \cite{Xie:2013gma}, holography \cite{Xie:2019qmw},  and etc. These constructions provide us with lots of interesting $\mathcal{N}=1$ 
SCFTs, and some properties of them can be found using the powerful string theoretical methods \cite{Witten:1997ep,Xie:2013rsa}.
 
 It is important to further enlarge the space of $\mathcal{N}=1$ SCFTs. In \cite{Maruyoshi:2018nod} all possible $\CN=1$ preserving deformations of $(A_1, A_2)$ and $(A_1, A_3)$ theories are worked out. Then it was  shown in \cite{Xie:2019aft} that one can derive a large class of new $\mathcal{N}=1$ SCFTs by turning on $\mathcal{N}=1$ preserving relevant deformations of $\mathcal{N}=2$ SCFT. The $\mathcal{N}=2$ SCFTs used in \cite{Xie:2019aft} are the so-called strongly coupled Argyres-Douglas theories which consist of a variety of available deformations, i.e. there are a large number of  operators with scaling dimension $\Delta({\cal O})<3$, which
 we can use to turn on $\mathcal{N}=1$ preserving deformation \footnote{See \cite{Benini:2009mz} for the construction of $\mathcal{N}=1$ SCFT by using marginal operators of $\mathcal{N}=2$ SCFT. }. This is in contrast with $\mathcal{N}=2$ Lagrangian theories or the theories considered in \cite{Benini:2009mz}, where the only Coulomb branch operators with dimension less than three is the 
 marginal operator.  The power of the construction studied in \cite{Xie:2019aft} is that one can learn interesting properties about the resulting $\mathcal{N}=1$ SCFTs using  results of $\mathcal{N}=2$ SCFTs,
 such as the central charges $a, c$, the scaling dimension of the chiral operators, etc. These properties are often not easy to compute 
 for $\mathcal{N}=1$ SCFTs. 

The purpose of this note is to further study  $\mathcal{N}=1$ SCFTs constructed in \cite{Xie:2019aft} by using the $\mathcal{N}=1$ superconformal index (index for short), which can be derived using known indices of 
parent $\mathcal{N}=2$ theories. The $\mathcal{N}=1$ index is quite useful  to learn the phase structure of IR $\mathcal{N}=1$ SCFT. The Schur index of a large class of $\mathcal{N}=2$ AD theories was found for \cite{Cordova:2015nma, Song:2017oew, Xie:2019zlb}, and recently even the full index has been found for some AD theories \cite{Maruyoshi:2016tqk, Maruyoshi:2016aim, Agarwal:2016pjo, Agarwal:2018ejn}. 
We will show that it is straightforward to get IR $\mathcal{N}=1$ index once the UV $\mathcal{N}=2$ index is known.

Once we determine the IR $\mathcal{N}=1$ index, we can use it to probe the phase structure of IR theory, and we find the following results:
\begin{enumerate}
\item If the IR theory is an interacting SCFT, one can get the information on the chiral ring from the single letter index. This confirms the conjecture proposed in \cite{Buican:2016hnq, Xie:2016hny} for the simplest case.
\item In many cases, the IR theory contains just free chirals. For example, if we turn on the deformation using the $\mathcal{N}=2$ Coulomb branch operators with the smallest scaling dimension of the $(A_1, A_{2N})$ theory, the resulting IR theory 
is just a free chiral! We found a large class of RG flows such that the IR $\mathcal{N}=1$ theories consist of just free chirals.
\item We also have the intermediate cases where the IR theory consists of free chirals plus an interacting theory. We find that often the IR theory may have accidental symmetry.
\end{enumerate}

This paper is organized as follows: in section \ref{sec:index} we review the $\mathcal{N}=1$ preserving deformation of $\mathcal{N}=2$ SCFT, and the way to derive $\mathcal{N}=1$ index from the index of the parent $\mathcal{N}=2$ theory. Section \ref{sec:chiralring} discusses information on $\mathcal{N}=1$ chiral ring extracted from the index; Section \ref{sec:freechiraldeform} discusses the case where IR theory is just free chirals; Section \ref{sec:otherPheno} discusses other interesting phenomenon such as the decoupled sector and accidental symmetry; finally
a conclusion is given in section \ref{sec:conc}.

\section{ $\CN=1$ index from $\CN=2$ index }
\label{sec:index}

\subsection{$\mathcal{N}=1$ SCFT from $\mathcal{N}=2$ SCFT}
We  first review some representation theory results of 4d $\mathcal{N}=2$ superconformal algebra. 
The bosonic symmetry group of a general $\mathcal{N}=2$ SCFT is $SO(2,4)\times SU(2)_R \times U(1)_r \times G_F$, where $SO(2,4)$ 
is the conformal group, $SU(2)_R\times U(1)_r$ is the $R$ symmetry group which exists for every $\mathcal{N}=2$ SCFT, and $G_F$ are other 
global symmetry groups commuting with the superconformal group which could be absent for some theories. A highest weight representation of the $\CN=2$ superconformal group is labelled by $|\Delta,R,r, j_1, j_2\rangle$, where $\Delta$ is
the scaling (conformal) dimension, $r$ is $U(1)_r$ charge, $R$ is $SU(2)_R$ spin, $j_1$ and $j_2$ are left and right spin. These states 
could also carry quantum numbers of flavor symmetry group $G_F$. Representation theory of $\mathcal{N}=2$ superconformal group in $3+1$d has been studied in \cite{Dolan:2002zh}.
and  short representations are completely classified in \cite{Dolan:2002zh}. Three short representations that we are interested in  are $\CE_{r,(0,0)}$ multiplets containing Coulomb branch operators, $\hat\CB_R$ multiplets containing
Higgs branch operators, and supercurrent multiplet $\hat\CC_{0,(0,0)}$. Their respective BPS conditions are listed below.
\begin{equation}
\begin{split}
&\text{Coulomb~branch operators}:~~~~{\bar{\cal E}}_{r,(0,0)},~~R=0,~~\Delta=r\footnote{We call $\bar\CE_{r(j_1,0)}$ chiral multiplet because its highest weight state is annihilated by $\tilde{Q}_{i\dot\alpha}$, and our $r$ is $-r$ in \cite{Gadde:2011uv} so that the BPS condition of $\CN=2$ chiral multiplet $\bar\CE_{r(j_1,0)}$ is $\Delta=r$}, \nonumber\\
&\text{Higgs~branch operators}:~~~~~~~~\hat{{\cal B}}_R,~~~~~~~r=j_1=j_2=0,~~\Delta=2R, \nonumber\\
&\text{Supercurrent}:~~~~~~~~~~~~~~~~~~~~~\hat{{\cal C}}_{0,(0,0)},~~r=R=0,~~\Delta=2. \nonumber\\
\end{split}
\end{equation}
For example, if the theory has nontrivial flavor symmetry, $\hat{{\cal B}}_1$ multiplets contains conserved currents for the flavor symmetry group $G_F$, and  transforms in the adjoint representation of $G_F$.
The $\mathcal{N}=1$ subalgebra is generated by the supercharge $Q_{1\alpha}$, and the corresponding generator of the $\CN=1$ $R$ symmetry is $\hat{R}_{\mathcal{N}=1}={2\over 3} \hat{r}+{4\over 3} \hat{I}_3$ \footnote{Here $\hat{r}$ is the generator for $\mathcal{N}=2$ $U(1)_r$ symmetry, and $\hat{I}_3$ is the generator of the Cartan subalgebra of Lie algebra associated with $SU(2)_R$ symmetry. In our convention a letter with hat means the generator and the same letter without hat means the eigenvalue of the generator.}. The other global 
symmetry group in $\mathcal{N}=1$  description is generated by $\hat{J}=2\hat{r}-2\hat{I}_3$ which commutes with the supercharge $Q_{1\alpha}$ \footnote{In our notation $(Q_{1\alpha}, Q_{2\alpha})$ 
are spin $SU(2)$ doublet with $I_3(Q_1)=-{1\over2}, I_3(Q_2)={1\over2}$, and $U(1)_r$ charges are  $R(Q_1)=R(Q_2)=-{1\over2}$.}.

One can engineer a large class of 4d $\mathcal{N}=2$ SCFTs by using either 6d $(2,0)$ construction \cite{Gaiotto:2009we, Xie:2012hs} or Type IIB string theory on three-fold singularity \cite{Xie:2015rpa}. The great advantage of these constructions are 
that one can naturally determine the Coulomb branch spectrum and the Higgs branch structure of these theories. 

The soft breaking terms of 4d $\mathcal{N}=2$ SCFT are classified in \cite{Xie:2019aft}. In this paper, we are interested in $\mathcal{N}=1$ preserving deformation by using Coulomb branch operators which form a freely generated chiral ring with finitely many generators obtained from the Seiberg-Witten (SW) geometry in $\CN=2$ SCFTs. 
The  $\mathcal{N}=2$ $\bar\CE_{r(0,0)}$ multiplet containing the Coulomb branch operator consists of 
three $\mathcal{N}=1$ chiral multiplets $({\cal O},\lambda_{\alpha}, S)$, and only multiplet ${\cal O}$ \footnote{When there is no ambiguity, we will use the same symbol to denote the highest weight primary operator, its corresponding multiplet and superfield to avoid clutter.}
 will give new $\mathcal{N}=1$ preserving deformations.
If we start with a $\bar{\cal E}_{r_0,(0,0)}$ multiplet with $U(1)_r$ charge $r_0$ of $\CN=2$ theory $\CT^{\CN=2}$,  we can use its buttom $\mathcal{N}=1$ chiral multiplet ${\cal O}_{\bar\CE_{r_0,(0,0)}}$ to deform our theory,
\begin{equation}
\label{eq:CoulombDef}
\delta S =\lambda \int d^2\theta {\cal O}_{\bar\CE_{r_0,(0,0)}}+c.c
\end{equation} 
The resulting IR $\mathcal{N}=1$ SCFT will be denoted as $\CT^{\CT^{\CN=2}}[\CO_{\bar\CE_{r_0,(0,0)}}]$ to indicate its $\CN=2$ parent theory and the deformation or as $\CT[\CO_{\bar\CE_{r_0,(0,0)}}]$ when the parent theory is apparent. 
The candidate generator of the $U(1)_R$ symmetry for the IR theory $\CT[\CO_{\bar\CE_{r_0,(0,0)}}]$ is 
\begin{equation}
\hat{R}_{IR}={2\over r_0}\hat{r}+({2-{2\over r_0}})\hat{I}_3.
\label{iru1}
\end{equation}
We have to emphasize that the IR theory $\CT[\CO_{\bar\CE_{r_0,(0,0)}}]$ might have accidental symmetry so that the correct generator of the $U(1)_R$ symmetry would be different from the above one, and one can find the true generator of the IR $U(1)_R$ symmetry using 
the $a$-maximization procedure \cite{Intriligator:2003jj}.

Some key properties of $\CT[\CO_{\bar\CE_{r_0,(0,0)}}]$  can be found as follows (consider only Coulomb type deformations \ref{eq:CoulombDef} and assume that the candidate generator  \ref{iru1} is the true generator of the IR $U(1)_R$ symmetry of $\CT[\CO_{\bar\CE_{r_0,(0,0)}}]$):
\begin{enumerate}
\item \textbf{Central charges}: One can compute the IR central charge as follows (central charges of $\mathcal{N}=1$ deformations of $(A_1,A_N) $ theories are also discussed in \cite{Giacomelli_2015}):
\begin{equation}
\begin{split}
&a_{N=1}=(a_{N=2}-c_{N=2})[{27\over16}x^3-{9\over4}x]+(2a_{N=2}-c_{N=2})[{27\over32}xy^2], \nonumber\\
&c_{N=1}=(a_{N=2}-c_{N=2})[{27\over16}x^3-{15\over4}x]+(2a_{N=2}-c_{N=2})[{27\over32}xy^2].
\end{split}
\label{ircentral}
\end{equation}
Here $x={2\over r_0},y=2-{2\over r_0}$. 
\item \textbf{Index}: The explicit form of $\mathcal{N}=2$ Schur index of many interesting theories is known \cite{Cordova:2015nma, Song:2017oew, Xie:2019zlb}, and this index is actually invariant under RG flow, and can be used to get some useful information of IR theory \cite{Buican:2016hnq}.
Details are discussed in the next subsection.
\item \textbf{Chiral ring}: The $\mathcal{N}=1$ chiral operator ${\cal O}_{\bar\CE_{r_0,(0,0)}}$ which is used to deform our theory satisfies a chiral ring relation ${\cal O}_{\bar\CE_{r_0,(0,0)}}=0$ in the IR theory \cite{Xie:2016hny,Buican:2016hnq}. 
\item \textbf{Chiral spectrum}: One can get some information of chiral operators of $\CT[\CO_{\bar\CE_{r_0,(0,0)}}]$ from parent $\mathcal{N}=2$ theory.  For an $\mathcal{N}=2$ chiral operator $\bar\CE_{r,(0,0)}$ with $U(1)_r$ charge $r$, the scaling dimension of its three $\mathcal{N}=1$ chiral multiplets after deformation using ${\cal O}_{\bar\CE_{r_0,(0,0)}}$ are
\begin{equation}
[{\cal O}_{\bar\CE_{r,(0,0)}}]={3r\over r_0},~~[\lambda_{\alpha}^{\bar\CE_{r,(0,0)}}]={{3\over2}r_0-3+3r\over r_0},~~~[S_{\bar\CE_{r,(0,0)}}]={3(r_0+r-2)\over r_0}.
\end{equation}
 If $2<r_0<3$, the minimal scaling dimension of a chiral scalar operator is $\Delta_{min}= {3r_{min}\over r_0}$;
if $1<r_0\leq 2$, we have $\Delta_{min}= {3(r_0+r_{min}-2)\over r_0}$. 
In particular, some of these chiral operators are relevant, and one 
can use them to further deform $\CT[\CO_{\bar\CE_{r_0,(0,0)}}]$ and flow to possibly new  $\mathcal{N}=1$ SCFTs, although these deformations would break the $\CN=1$ $U(1)_R$ symmetry, and we have little to say about IR theory.  The flavor symmetry  of UV $\mathcal{N}=2$ theory
is not broken by the Coulomb branch type $\mathcal{N}=1$ preserving deformations \ref{eq:CoulombDef}, so we do know the existence of $\hat{\cal C}_{(0,0)}$ type multiplets in $\CT[\CO_{\bar\CE_{r_0,(0,0)}}]$ from the $\CN=2$ multiplets containing the flavor symmetry current of the UV theory.
If $\mathcal{N}=2$ theory also have $\hat{B}_R$ type operators, one also get chiral operators $X_R$ in $\CT[\CO_{\bar\CE_{r_0,(0,0)}}]$ with scaling dimension ${3\over 2}(2-{2\over r_0})R$. If there is operator with $R=1$, then the operator with smallest scaling dimension in the IR theory $\CT[\CO_{\bar\CE_{r_0,(0,0)}}]$ would be $X_1$ with 
scaling dimension $3-{3\over r_0}$ for $ 1<r_0 \leq 2$. 

\item \textbf{Exact marginal deformations}: Firstly If the UV $\mathcal{N}=2$ theory has a multiple number of Coulomb branch operators with the same $U(1)_r$ charge $r_0$, 
 the IR theory $\CT[\CO_{\CE_{r_0,(0,0)}}]$ might have exact marginal deformations. Secondly if the UV $\mathcal{N}=2$ theory has a $\mathcal{N}=2$ exact marginal operator,  the $S$ component of it would be exact marginal in the IR $\mathcal{N}=1$ SCFT. 
Finally if $r_0=2$, then the $\hat{{\cal B}}_2$ type operator might also give exact marginal deformations of $\CT[\CO_{\bar\CE_{r_0,(0,0)}}]$. 

\item \textbf{Inherited $\mathcal{N}=1$ duality}: If the 4d $\mathcal{N}=2$ theory has an exact marginal deformation and has different duality frames,   the IR theory would also have different duality frames. There are many $\mathcal{N}=2$ theories whose duality frames are known \cite{Xie:2016uqq,Xie:2017vaf,Xie:2017aqx}, and using the $\mathcal{N}=1$ preserving deformations, we 
get a large class of new Seiberg-like duality for $\mathcal{N}=1$ SCFTs.
\end{enumerate}

\subsection{The $\CN=2$ superconformal index and its Schur limit}
\label{sec:fullindex}
As we discussed in last subsection, one can get the index of IR $\mathcal{N}=1$ SCFT by using the known result of $\mathcal{N}=2$ index. Here we review the relation between the indexes of UV $\mathcal{N}=2$ and IR $\mathcal{N}=1$ theory.
The $\CN=2$ superconformal index \cite{Kinney:2005ej,Romelsberger:2005eg} with respect to the $SU(2,2|2)$ supercharge $Q_{1\dot{-}}$ is defined as 
\begin{equation}
\label{eq:def:N2index}
\CI^{\CN=2}=\Tr(-1)^Fp^{j_1+j_2+r}q^{-j_1+j_2+r}t^{R-r}e^{-\beta\delta_{1\dot{-}}},
\end{equation}
where $\delta_{1\dot{-}}=2\{\tilde{Q}_{1\dot{-}},\tilde{Q}_{1\dot{-}}^\dag\}=\Delta-2j_2-2R-r$. $\Delta$, $j_1$, $j_2$, $R$, $r$ are conformal dimension, two Lorentz spins, $SU(2)_R$ and $U(1)_r$ respectively which are Cartans of the $SU(2,2|2)$. Notice that our $r$ is $-r$ in \cite{Gadde:2011uv} so that the BPS condition of $\CN=2$ chiral multiplet $\bar\CE_{r(j_1,0)}$ is $\Delta=r$\footnote{We call $\bar\CE_{r(j_1,0)}$ chiral multiplet because its highest weight state is annihilated by $\tilde{Q}_{i\dot\alpha}$.}. Supersymmetry ensures that the trace is only over BPS states satisfying $\delta_{1\dot{-}}=0$, i.e. state satisfying $\Delta-2j_2-2R-r=0$. Note that the highest weight primary operators of $\hat{\CB}_R$ and $\bar\CE_{r(j_1,0)}$ multiplets and other deformations used in this paper preserve $\tilde{Q}_{1\dot-}$ supercharge, hence they all contribute to the index. Information on $\CN=2$ BPS multiplets and their superconformal indices are summarized in appendix \ref{app:sec:BPSmultiplets}.

Unfortunately the full superconformal indices of a large portion of $\CN=2$ theories are  unknown yet, however, in many such theories certain limits of the full indices can still be computed exactly\cite{Cordova:2015nma,Song:2015wta,  Song:2016yfd,Buican:2017uka, Song:2017oew, Xie:2019zlb}. One of such example is the Schur index which is the full index at $t=q$ limit. In this limit, only BPS operators in Schur sector which are annihilated by both $Q_{1+}$ and $\tilde{Q}_{1\dot{-}}$  (i.e. operators satisfy $\delta_{1+}=\Delta-2j_1-2R+r=0$ and $\delta_{1\dot{-}}=\Delta-2j_2-2R-r=0$) contribute to the index, and the dependence on $p$ is automatically dropped out. The $\bar\CE_r=\bar\CE_{r(0,0)}$ multiplets containing Coulomb branch operators do not contribute in the Schur limit, while the $\hat{\CB}_R$ multiplets still contribute. Multiplets which contribute to the Schur index are also summarized in appendix \ref{app:sec:BPSmultiplets}.

\subsubsection{The $\CN=1$ index from $\CN=2$ index}
\label{sec:Schur}

Once we deform the theory with $\lambda \int d^2\theta {\cal O}_{\bar\CE_{r_0,(0,0)}}+c.c$ which preserves the $\tilde{Q}_{1\dot-}$ supercharge, the infrared theory has only $\CN=1$ superconformal symmetry. One can still define the index with respect to $\tilde{Q}_{1\dot-}$, but only the IR $U(1)_R$ symmetry generated by $\hat{R}_{IR}=\frac{2}{r_0}\hat{r}+\left(2-\frac{2}{r_0}\right)\hat{I}_3$ is preserved in the IR fixed point.  If we substitute $t=(pq)^{1-\frac{1}{r_0}}$ in the definition $\CN=2$ index \ref{eq:def:N2index}, we get
\begin{equation}
\Tr(-1)^Fp^{j_1+j_2+\frac{R_{IR}}{2}}q^{-j_1+j_2+\frac{R_{IR}}{2}}e^{-\beta\delta_{1\dot-}},
\end{equation}
which is exactly the definition of $\CN=1$ superconformal index \cite{Kinney:2005ej,Romelsberger:2005eg} with respect to $Q_{1\dot-}$. This substitution is a generalization of the procedure in \cite{Gadde:2010en}.

Our strategy to get the $\CN=1$ index of the IR theory is therefore working out the $\CN=2$ index for the theory we would like to deform first, then pick the $\CN=1$ preserving deformation and make the corresponding substitution in $t$ fugacity and obtain the $\CN=1$ index of the IR theory. Once we get the $\CN=1$ index, we can obtain the information on the IR theory.

In \cite{Buican:2016hnq}  authors  proposed another limit to deal with theories of which only the Schur index is known. They took the limit $t=q$ and $p=q^{\frac{1}{r_0-1}}$ in the full index \ref{eq:def:N2index}, 
\begin{equation}
\label{eq:N1Schurindex}
\CI_{S}^{\CN=1}=\CI^{\CN=2}\left(p=q^{\frac{1}{r_0-1}},q,t=q\right)=\Tr (-1)^Fq^{-\frac{r_0-2}{r_0-1}j_1+\frac{r_0}{r_0-1}j_2+\frac{r_0}{2(r_0-1)}R_{IR}}e^{-\beta\delta_{1\dot-}}.
\end{equation}
Since the index under $t=q$ limit (which is the Schur limit) is independent of $p$, $\CI^{\CN=1}_S$ should be the \textbf{same} as the Schur index of the UV $\CN=2$ theory. On the other hand, since the deformation preserve the supercharge $Q_{1-}$, $\CI^{\CN=1}_S$ is a good index for the IR $\CN=1$ theory albeit the fugacity $q$ labels an unconventional combination of the Cartans of the $\CN=1$ superconformal algebra. Although the $\bar\CE_{r,(0,0)}$ type multiplets which represent Coulomb branch operators do not contribute to the Schur index, some valuable information on their relations in the deformed theory can still be extracted from $\CI^{\CN=1}_{S}(q)=\CI^{\CN=2}_{Schur}(q)$.

\section{Interacting $\mathcal{N}=1$ SCFTs and their chiral ring}
\label{sec:chiralring}

In this section, we will use the known result of full $\mathcal{N}=2$ index to get the full index of the IR $\mathcal{N}=1$ theory after deformation. 
The full index for  $\mathcal{N}=2$ $(A_1, A_k)$ and $(A_1,D_k)$ theories
have been worked in  \cite{Maruyoshi:2016tqk, Maruyoshi:2016aim, Agarwal:2016pjo}, together with vanishing OPEs for a large class of AD theories discussed in \cite{Agarwal:2018zqi}. Using the proposal in last section, we can find 
the full index for the $\mathcal{N}=1$ SCFT derived from turning on the $\mathcal{N}=1$ preserving relevant deformation, then one can further derive some part of $\mathcal{N}=1$ chiral ring. 

\subsection{$(A_1, A_{2N})$ results}
\label{sec:A1A2N}

\subsubsection{$\CN=2$ index}


The full index of $(A_1, A_{2N})$ AD theories are given in \cite{Maruyoshi:2016tqk, Maruyoshi:2016aim, Agarwal:2016pjo} as an integration over elliptic Gamma functions
\begin{equation}
\label{eq:A1A2NfullN2index}
\begin{split}
\CI^{(A_1, A_{2N})}_{\CN=2}
=&\left[\prod_{i=1}^N\frac{\Gamma\left(\left(\frac{pq}{t}\right)^{\frac{2(N+i+1)}{2N+3}}\right)}{\Gamma\left(\left(\frac{pq}{t}\right)^{\frac{2i}{2N+3}}\right)}\right]\Gamma\left(\left(\frac{pq}{t}\right)^{\frac{1}{2N+3}}\right)^N\\
&\times\frac{\kappa^N}{2^NN!}\oint[d\bfz]\prod_{\alpha\in\Delta}\frac{\Gamma\left(\bfz^\alpha\left(\frac{pq}{t}\right)^\frac{1}{2N+3}\right)}{\Gamma(\bfz^\alpha)}
\prod_{w\in R}\Gamma\left(\bfz^w\left(\frac{pq}{t}\right)^\frac{N+1}{2N+3}t^\half\right)\Gamma\left(\bfz^w\left(\frac{pq}{t}\right)^\frac{-N}{2N+3}t^\half\right),
\end{split}
\end{equation}
where $[d\bfz]=\prod_{i=1}^N\frac{dz_i}{2\pi iz_i}$, $\Delta$ is the set of roots of $Sp(N)$ and $R$ is the set of weights of the fundamental representation of $Sp(N)$. The properties of elliptic Gamma function $\Gamma(z;p,q)$ are summarized in appendix \ref{app:sec:eGamma}.

To better read off the generators and relations, we write the indices in the form of the Plethystic exponential of single letter indices \begin{equation}\CI=\pe{\CI^{s.l.}},\end{equation} where $\pe{f(a,b,c,\cdots)}$ is the Plethystic exponential 
\begin{equation}
\pe{f(a,b,c,\cdots)}=\exp\left(\sum_{n=1}^\infty\frac{1}{n}f(a^n,b^n,c^n,\cdots)\right).
\end{equation}
$\CI^{s.l.}$ contains all basic building blocks and their relations in the spectrum, then the Plethystic exponential automatically take care of the products of these operators according to correct statistics.

With the integral formula, the expansions of single letter indices up to the first few orders when $N=1,2,3$ are 
\begin{equation}
\CI^{s.l.}_{(A_1, A_2)}=\CI_{\bar{\CE}_{\frac{6}{5}}}+\CI_{\hat{\CC}_{0(0,0)}}-\CI_{\bar{\CB}_{1,\frac{7}{5}(0,0)}}+\cdots,
\end{equation}
\begin{equation}
\CI^{s.l.}_{(A_1, A_4)}=\CI_{\bar{\CE}_{\frac{8}{7}}}+\CI_{\bar{\CE}_{\frac{10}{7}}}+\CI_{\hat{\CC}_{0(0,0)}}-\CI_{\bar{\CB}_{1,\frac{9}{7}(0,0)}}-\CI_{\bar{\CB}_{1,\frac{11}{7}(0,0)}}-\CI_{\bar{\CB}_{1,\frac{13}{7}(0,0)}}+\cdots,
\end{equation}
and
\begin{equation}
\begin{split}
\CI^{s.l.}_{(A_1, A_6)}=&\CI_{\bar{\CE}_{\frac{10}{9}}}+\CI_{\bar{\CE}_{\frac{4}{3}}}+\CI_{\bar{\CE}_{\frac{14}{9}}}+\CI_{\hat{\CC}_{0,(0,0)}}\\
&-\CI_{\bar{\CB}_{1,\frac{11}{9}(0,0)}}-\CI_{\bar{\CB}_{1,\frac{13}{9}(0,0)}}-\CI_{\bar{\CB}_{1,\frac{5}{3}(0,0)}}-\CI_{\bar{\CB}_{1,\frac{17}{9}(0,0)}}-\CI_{\bar{\CB}_{1,\frac{19}{9}(0,0)}}+\cdots.
\end{split}
\end{equation}
Here $\CI_{\CM}$ is the index of the $\CN=2$ multiplet $\CM$ with $\CM$ being $\CI_{\hat{\CC}_{0(0,0)}}$, $\bar{\CE}_r$\footnote{We denote $\bar\CE_{r(0,0)}$  as $\bar\CE_{r}$ for short.}, or $\bar{\CB}_{1,r(0,0)}$,
\begin{equation}
\begin{split}
\CI_{\hat{\CC}_{0(0,0)}}&=\frac{(t-pq)(p+q-pq/t)}{(1-p)(1-q)},\\
\CI_{\bar{\CE}_r}&=\left(\frac{pq}{t}\right)^r\frac{\left(1-\frac{t}{p}\right)\left(1-\frac{t}{q}\right)}{(1-p)(1-q)},\\
\CI_{\bar{\CB}_{1,r(0,0)}}&=\left(\frac{pq}{t}\right)^r\frac{(t-pq)\left(1-\frac{t}{p}\right)\left(1-\frac{t}{q}\right)}{(1-p)(1-q)}.
\end{split}
\end{equation}
And in general we conjecture the single letter index of $(A_1, A_{2N})$ theory always starts with
\begin{equation}
\label{eq:N2indexA1A2N}
\CI^{s.l.}_{(A_1, A_{2N})}=\sum_{i=1}^N\CI_{\bar{\CE}_{\frac{2N+2i+2}{2N+3}}}+\CI_{\hat{\CC}_{0(0,0)}}-\sum_{i=1}^{2N-1}\CI_{\bar{\CB}_{1,\frac{2N+2i+3}{2N+3}(0,0)}}+\cdots.
\end{equation}

From the single letter index one immediately sees that at lower conformal dimension, the spectrum of $(A_1, A_{2N})$ AD theories contains the stress tensor multiplet $\hat{\CC}_{0(0,0)}$ and $N$ $\bar{\CE}_{r_i}$ type multiplets with $r_i=\frac{2N+2i+2}{2N+3}$ and $1\leq i\leq N$, which are $N$ generators of the Coulomb branch chiral ring of the theory. One can also read off relations of these operators as well. Superconformal representation theory tells us that the tensor product of two $\bar{\CE}$-type multiplets $\bar{\CE}_{r_1}$ and $\bar{\CE}_{r_2}$ gives
\begin{equation}
\bar{\CE}_{r_1}\otimes \bar{\CE}_{r_2}\simeq\bar{\CE}_{r_1+r_2}\oplus\bar{\CB}_{1,r_1+r_2-1(0,0)}+\cdots.
\end{equation}
In terms of index, this means
\begin{equation}
\CI_{\bar{\CE}_{r_1}}\times\CI_{\bar{\CE}_{r_2}}=\CI_{\bar{\CE}_{r_1+r_2}}+\CI_{\bar{\CB}_{1,r_1+r_2-1(0,0)}}+\cdots,
\end{equation}
In the full index, if one expand in terms of conformal dimension, one has
\begin{equation}
\begin{split}
\pe{\CI_{\bar\CE_{r_1}}+\CI_{\bar\CE_{r_2}}}&=1+\CI_{\bar\CE_{r_1}}+\CI_{\bar\CE_{r_2}}+\CI_{\bar\CE_{r_1}\otimes\bar\CE_{r_2}}+\cdots,\\
&=1+\CI_{\bar\CE_{r_1}}+\CI_{\bar\CE_{r_2}}+\CI_{\bar\CE_{r_1+r_2}}+\CI_{\bar\CB_{1,r_1+r_2-1(0,0)}}+\cdots,
\end{split}
\end{equation}
then a $-\CI_{\bar\CB_{1,r_1+r_2-1(0,0)}}$ term in the single letter index will cancel the $\CI_{\bar\CB_{1,r_1+r_2-1(0,0)}}$ contribution from tensor product $\bar\CE_{r_1}\otimes\bar\CE_{r_2}$  in the full index, leaving only
\begin{equation}
\pe{\CI_{\bar\CE_{r_1}}+\CI_{\bar\CE_{r_2}}-\CI_{\bar\CB_{1,r_1+r_2-1(0,0)}}}=1+\CI_{\bar\CE_{r_1}}+\CI_{\bar\CE_{r_2}}+\CI_{\bar\CE_{r_1+r_2}}+\cdots,
\end{equation}
hence the $-\CI_{\bar\CB_{1,r_1+r_2-1(0,0)}}$ term in the single letter index implies the relation of operators
\begin{center}
\begin{tcolorbox}
\begin{equation}
\left(\bar\CE_{r_1}\otimes\bar\CE_{r_2}\right)_{\bar\CB_{1,r_1+r_2-1(0,0)}}=0,
\end{equation}
\end{tcolorbox}\end{center}
where $(\cdots)_{\CM}$ means picking only the $\CM$ multiplet appeared in the $\cdots$. It is important to notice that each distinct value of $r_i+r_j-1$ appears only once in the $-\sum_{i=1}^{2N-1}\CI_{\bar{\CB}_{1,\frac{2N+2i+3}{2N+3}(0,0)}}$ term of the single letter index \ref{eq:N2indexA1A2N} when $r_i$ and $r_j$ run over all possible $U(1)_r$ charges of the Coulomb branch operators. Therefore if there are two pairs of chiral multiplets $(\bar\CE_{r_i},\bar\CE_{r_j})$ and $(\bar\CE_{r_k},\bar\CE_{r_l})$ satisfying $r_i+r_j=r_k+r_l$, there is only one relation
\begin{equation}
\left(\bar\CE_{r_i}\otimes\bar\CE_{r_j}\oplus\bar\CE_{r_k}\otimes\bar\CE_{r_l}\right)_{\bar\CB_{1,r_i+r_j-1(0,0)}}=0.
\end{equation}
The same is true for multiple pairs of chiral multiplets with the same total $U(1)_r$ charge, in the end there are $2N-1$ relations for $\bar\CE_r$ type multiplets of $(A_1, A_{2N})$ for $2N-1$ $-\CI_{\bar\CB_{1,r_1+r_2-1(0,0)}}$ terms in the single letter index,
\begin{center}
\begin{tcolorbox}\begin{equation}
\sum_{r'\in\CA, r+1-r'\in\CA}\left(\bar\CE_{r'}\otimes\bar\CE_{r+1-r'}\right)_{\bar\CB_{1,r(0,0)}}=0,\,\,\,\,\mathrm{for}\,\ r=\frac{2N+5}{2N+3},\,\frac{2N+7}{2N+3},\cdots,\frac{4N+1}{2N+3},
\label{eq:N2relationA1A2N}
\end{equation}\end{tcolorbox}\end{center}
where $\CA=\{\frac{2N+4}{2N+3},\frac{2N+6}{2N+3},\cdots,\frac{4N+2}{2N+3}\}$ is the set of $U(1)_r$ charges of $(A_1, A_{2N})$ theories.

\subsubsection{$\CN=1$ index and chiral ring}
\label{sec:N1deformA1A2N}

One $\bar\CE_r$ multiplet splits into three $\CN=1$ chiral multiplets whose primary operators are $\CO_{\bar\CE_r}$, $\lambda_{\bar\CE_r,\alpha}$ and $S_{\bar\CE_r}$ respectively\footnote{Again we use the same symbol to denote both the multiplet and its primary operator to avoid complications in notations}. Let $\{\bar\CE_{r_1},\cdots,\bar\CE_{r_l}\}$ be generators of the Coulomb branch chiral ring of an AD theory, then the number of generators $l$ is also called the rank of this AD theory. In the following, we will further denote the primary operators of these multiplets $\{u_i|1\leq i\leq l\}$ to separate them from $\CO_r$ of a generic $\bar\CE_{r(0,0)}$, and order them from smallest to largest conformal dimension (i.e. $u_1$ is the operator with the smallest conformal dimension). According to the previous section, this theory admits relevant $\CN=1$ preserving deformation $\lambda \int d^2\theta {\cal O}_{\bar\CE_{r_0,(0,0)}}+c.c$ where $\CO$ can be chosen from $\{u_i,u_i u_j|1\leq i\leq j\leq l\}$ or any linear combinations of operators in this set with the same quantum number because all generators of the Coulomb branch chiral ring has scaling dimension greater than $1$ and relevant deformation requires the dimension of $\CO_{\bar\CE_{r_0(0,0)}}$ smaller than $3$. 

To get the index of the IR theory, we first substitute $t=(pq)^{1-\frac{1}{r_0}}$ in $\CN=2$ single letter indices, then one can read off the spectrum and constraints at lower dimension, together with other interesting informations on the IR theory. 

%

{\bf $\CN=2$ multiplets as $\CN=1$ multiplets:} We fist study how various  $\CN=1$ multiplets arises from $\CN=2$ multiplets under the deformation $\lambda \int d^2\theta {\cal O}_{\bar\CE_{r_0,(0,0)}}+c.c$. A summary of $\CN=1$ BPS multiplets and their indices are given in appendix \ref{app:sec:N1BPS}. 

The $\CN=2$ $\bar\CE_r$ type multiplets decompose into three $\CN=1$ chiral multiplets, and its index of $\CI_{\bar\CE_{r}}$  under the deformation is
\begin{equation}
\begin{split}
\CI_{\bar\CE_{r}}\left(p, q, t=(pq)^{1-\frac{1}{r_0}}\right)=&\frac{(pq)^{\frac{r}{r_0}}}{(1-p)(1-q)}-\frac{(pq)^{\frac{r-1}{r_0}+\frac{1}{2}}\left(\sqrt{\frac{p}{q}}+\sqrt{\frac{q}{p}}\right)}{(1-p)(1-q)}+\frac{(pq)^\frac{r+r_0-2}{r_0}}{(1-p)(1-q)}\\
=&\CI_{\bar\CB_{\frac{2r}{r_0}(0,0)}}+\CI_{\bar\CB_{\frac{2r+r_0-2}{r_0}(\half,0)}}+\CI_{\bar\CB_{\frac{2r+2r_0-4}{r_0}(0,0)}},
\end{split}
\end{equation}
where $\CI_{\bar\CB_{r(j_1,0)}}=(-1)^{2j_1}\frac{(pq)^{\frac{r}{2}}\chi_{j_1}\left(\sqrt{\frac{p}{q}}\right)}{(1-p)(1-q)}$ is the index of chiral multiplet $\bar\CB_{r(j_1,0)}$\footnote{We call $\bar\CB_{r(j_1,0)}$ an $\CN=1$ chiral multiplet because its highest weight state is annihilated by $\tilde{Q}_{\dot\alpha}$.} and $\chi_j$ is the character of the spin-$j$ representation of $SU(2)$. Therefore an $\bar\CE_{r_i}$ type multiplet after deformation by $\CO$ decomposes into chiral multiplets $\CO_i$, $\lambda_{i,\alpha}$, $S_i$ with $\CN=1$ $R_{IR}$-charge $\frac{2r}{r_0}$, $1+\frac{2(r-1)}{r_0}$ and $2+\frac{2(r-2)}{r_0}$ respectively.

The index of $\hat{\CC}_{0(0,0)}$ multiplet under the deformation is
\begin{equation}
\begin{split}
\CI_{\hat{\CC}_{0(0,0)}}\left(p, q, (pq)^{1-\frac{1}{r_0}}\right)
=&\frac{-pq - pq(p+q) + (pq)^{\frac{3}{2}-\frac{1}{r_0}}\left(\sqrt{\frac{p}{q}}+\sqrt{\frac{q}{p}}\right)+(pq)^{1+\frac{1}{r_0}}}{(1-p)(1-q)}\\
=&\CI_{\hat{\CC}_{(0,0)}}+\CI_{\hat{\CC}_{(\half,\half)}}+\CI_{\bar\CC_{\frac{2}{r_0}-1(0,\half)}}-\CI_{\bar\CB_{3-\frac{2}{r_0}(\half,0)}}.
\end{split}
\end{equation}
Here $\hat{\CC}_{(\half,\half)}$ is the $\CN=1$ stress tensor multiplet, $\bar\CC_{\frac{2}{r_0}-1(0,\half)}$ is a $1/4$-BPS multiplet. $\bar\CB_{3-\frac{2}{r_0}(\half,0)}$ comes from the equations of motion in $\hat{\CC}_{0(0,0)}$ multiplet. Actually the full $\hat{C}_{0(0,0)}$ multiplet decomposes as
\begin{equation}
\hat{C}_{0(0,0)}\simeq\hat{\CC}_{(0,0)}\oplus\hat{\CC}_{(\half,\half)}\oplus \CC_{1-\frac{2}{r_0}(\half,0)}\oplus\bar{\CC}_{-1+\frac{2}{r_0}(0,\half)}
\end{equation}
in terms of $\CN=1$ multiplets.

The index of $\bar\CB_{1,r(0,0)}$ multiplet under the deformation is
\begin{equation}
\label{eq:N2barBsplit}
\begin{split}
&\CI_{\bar\CB_{1,r(0,0)}}\left(p,q,(pq)^{1-\frac{1}{r_0}}\right)\\
=&\frac{(pq)^{1+\frac{r-1}{r_0}}}{(1-p)(1-q)}-\frac{(pq)^{\frac{3}{2}+\frac{r-2}{r_0}}\left(\sqrt{\frac{p}{q}}+\sqrt{\frac{q}{p}}\right)}{(1-p)(1-q)}+\frac{(pq)^{2+\frac{r-3}{r_0}}}{(1-p)(1-q)}\\
&-\frac{(pq)^{1+\frac{r}{r_0}}}{(1-p)(1-q)}+\frac{(pq)^{\frac{3}{2}+\frac{r-1}{r_0}}\left(\sqrt{\frac{p}{q}}+\sqrt{\frac{q}{p}}\right)}{(1-p)(1-q)}-\frac{(pq)^{2+\frac{r-2}{r_0}}}{(1-p)(1-q)}\\
=&\CI_{\bar\CB_{2+\frac{2r-2}{r_0}(0,0)}}+\CI_{\bar\CB_{3+\frac{2r-4}{r_0}(\half,0)}}
+\CI_{\bar\CB_{4+\frac{2r-6}{r_0}(0,0)}}+\CI_{\bar\CC_{\frac{2r}{r_0}(0,0)}}
+\CI_{\bar\CC_{1+\frac{2r-2}{r_0}(\half,0)}}+\CI_{\bar\CC_{2+\frac{2r-4}{r_0}(0,0)}}.
\end{split}
\end{equation}
$\bar\CB_{1,r(0,0)}$ contains also $\CN=1$ multiplets which do not contribute to the index, and we will not write the full decomposition.

{\bf Chiral ring relations for general deformation:}
Now we deform $(A_1, A_{2N})$ theories with $\CN=1$ deformation discussed above, then the $\CN=1$ single letter index of $\CT^{(A_1, A_{2N})}[\CO_{\bar\CE_{r_0(0,0)}}]$ theory   is
\begin{equation}
\label{eq:slN1A1A2N}
\begin{split}
&\CI^{\CN=1,s.l.}_{(A_1, A_{2N})}=\CI^{s.l.}_{(A_1, A_{2N})}\left(p,q,(pq)^{1-\frac{1}{r_0}}\right)\\
=&\sum_{i=1}^N\left(\CI_{\bar\CB_{\frac{4(N+i+1)}{(2N+3)r_0}(0,0)}}+\CI_{\bar\CB_{\frac{2(2i-1)}{(2N+3)r_0}+1(\half,0)}}+\CI_{\bar\CB_{\frac{4(-N+i-2)}{(2N+3)r_0}+2(0,0)}}\right)\\
&+\left(\CI_{\hat{\CC}_{(0,0)}}+\CI_{\hat{\CC}_{(\half,\half)}}+\CI_{\bar\CC_{1-\frac{2}{r_0}(\half,0)}}-\CI_{\bar\CB_{3-\frac{2}{r_0}(\half,0)}}\right)
\\
&-\sum_{i=1}^{2N-1}\left(\CI_{\bar\CB_{\frac{4i}{(2N+3)r_0}+2(0,0)}}+\CI_{\bar\CB_{\frac{2(-2N+2i-3)}{(2N+3)r_0}+3(\half,0)}}
+\CI_{\bar\CB_{\frac{4(-2N+i-3)}{(2N+3)r_0}+4(0,0)}}\right)\\
&
-\sum_{i=1}^{2N-1}\left(\CI_{\bar\CC_{\frac{2(2N+2i+3)}{(2N+3)r_0}(0,0)}}+\CI_{\bar\CC_{\frac{4i}{(2N+3)r_0}+1(\half,0)}}+\CI_{\bar\CC_{\frac{2(-2N+2i-3)}{(2N+3)r_0}+2(0,0)}}\right)+\cdots.
\end{split}
\end{equation}
The first line on the RHS of the equality comes from $\CI_{\bar\CE_{r}}$'s , the second line comes from $\CI_{\hat\CC_{0(0,0)}}$, and the third and forth lines come from $-\CI_{\bar\CB_{1,r(0,0)}}$'s.

From the $\CN=1$ single letter index \ref{eq:slN1A1A2N} one can immediately read off the generators of the chiral ring, which are
\begin{center}
\begin{tcolorbox}
\begin{itemize}
\item {\bf Generators:} $u_i$, $\lambda_{i,\alpha}$ and $S_i$ for $1\leq i\leq N$, 
\end{itemize}
\end{tcolorbox}
\end{center}
and their $R_{IR}$-charges are $\frac{4(N+i+1)}{(2N+3)r_0}$, $\frac{2(2i-1)}{(2N+3)r_0}+1$, and $\frac{4(-N+i-2)}{(2N+3)r_0}+2$ respectively. $\hat\CC_{(\half,\half)}$ is the $\CN=1$ stress tensor multiplet. And the theory has an extra $\bar\CC_{1-\frac{2}{r_0}(\half,0)}$ from the $\CN=2$ stress tensor multiplet.

Now we analyze the chiral ring relations from the index \ref{eq:slN1A1A2N}. The operator $\CO_{\bar\CE_{r_0(0,0)}}$ we used to deform the theory is the bottom component of a $\bar\CB_{2(0,0)}$ multiplet in the IR. If it is a single trace operator $\CO_{\bar\CE_{r_0(0,0)}}=u_i$, its index $\CI_{\bar\CB_{2(0,0)}}$ appears as part of the single letter index \ref{eq:slN1A1A2N}. If it is a composite operator like $\CO_{\bar\CE_{r_0(0,0)}}=u_i u_j+\cdots$, its index $\CI_{\bar\CB_{2(0,0)}}$ appears in the full index from terms like $\CI_{\bar\CB_{r_i}}\CI_{\bar\CB_{r_j}}$ with $r_i$ and $r_j$ are $R_{IR}$ of $u_i$ and $u_j$ respectively. Notice there is also the identity
\begin{equation}
\CI_{\bar\CB_{2(0,0)}}+\CI_{\hat\CC_{(0,0)}}=0,
\end{equation}
so in the full index the contribution from $\CO_{\bar\CE_{r_0(0,0)}}$ cancels with the contribution from the $U(1)$ current multiplet $\hat\CC_{(0,0)}$ from the $\CN=2$ stress tensor multiplet, leading to the relation $\CO_{\bar\CE_{r_0(0,0)}}=0$. To understand this relation, recall the recombination rule \ref{app:eq:N1recombine} of the $\CN=1$ multiplets
\begin{equation}
\CA^2_{0(0,0)}\simeq\hat\CC_{(0,0)}\oplus\CB_{-2(0,0)}\oplus\bar\CB_{2(0,0)}.
\end{equation}
Hence the marginal operator $\CO_{\bar\CE_{r_0(0,0)}}$ recombines with the conserved current multiplet $\hat\CC_{(0,0)}$, and fails to be exact marginal. Acting $Q_{1\alpha}$ on $\CO_{\bar\CE_{r_0(0,0)}}$ leads to another $\CN=1$ chiral multiplet $\bar\CB_{3-\frac{2}{r_0}(\half,0)}$\footnote{We pick $\CN=1$ superconformal subalgebra of the $\CN=2$ SCA containing $Q_{2\alpha}$ and $\tilde{Q}_{1\dot{\alpha}}$.}. However, the $-\CI_{\bar\CB_{3-\frac{2}{r_0}(\half,0)}}$ term in the single letter index gives \ref{eq:slN1A1A2N} a relation $Q_{1\alpha}\CO_{\bar\CE_{r_0(0,0)}}=0$. 
All in all, the chiral ring relation coming from $\CN=2$ stress tensor multiplet $\CI_{\hat{\CC}_{0(0,0)}}$ are
\begin{center}
\begin{tcolorbox}
\begin{itemize}
\item {\bf Relation 1:} $\CO_{\bar\CE_{r_0(0,0)}}=0$,
\item {\bf Relation 2:} $Q_{1\alpha}\CO_{\bar\CE_{r_0(0,0)}}=0$.
\end{itemize}
\end{tcolorbox}
\end{center}
These two relations depend on the explicit form of $\CO_{\bar\CE_{r_0(0,0)}}$.

There are also chiral ring relations from the $\CN=2$ relations \ref{eq:N2relationA1A2N}. These relations are represented by $-\CI_{\bar\CB_{1,r(0,0)}}$ terms in the $\CN=2$ single letter index, and becomes constraints on $\bar\CB$ and $\bar\CC$ type multiplets as show in \ref{eq:N2barBsplit}. By matching the quantum number we can see the third line of \ref{eq:slN1A1A2N} gives the chiral ring relations
\begin{center}
\begin{tcolorbox}\begin{itemize}
\item {\bf Relation 3:} $\lambda_{i,\alpha}\lambda^\alpha_j+u_iS_j+u_jS_i+\cdots=0$,
\item {\bf Relation 4:} $S_i\lambda_{j,\alpha}+S_j\lambda_{i,\alpha}+\cdots=0$,
\item {\bf Relation 5:} $S_iS_j+\cdots=0$,
\end{itemize}
\end{tcolorbox}
\end{center}
where $\cdots$ means any other \textbf{possible} chiral operators with the same $R_{IR}$ charge. In certain cases, terms in $\cdots$ are necessary because of constraints from bootstrap method\cite{Poland:2015mta}. Each relation sums over all pairs of $(i,j)$ with $i+j$ remains the same, and the sum of $i+j$ takes all integers between $2$ and $2N$. The form of relation 3, 4 and 5 is universal and does not depend on the explicit form of $\CO_{\bar\CE_{r_0(0,0)}}$.

The last line of \ref{eq:slN1A1A2N} also gives relations
\begin{equation}
\begin{split}
&Q_{2}^\alpha\left(u_i\lambda_{j\alpha}+u_j\lambda_{i,\alpha}+\cdots\right),\\
&Q_{2\alpha}\left(\lambda_{i,\beta}\lambda^\beta+u_iS_j+u_jS_i+\cdots\right)=0,\\
&Q_{2}^\alpha\left(S_i\lambda_{j,\alpha}+S_j\lambda_{i,\alpha}+\cdots\right)=0,\\
\end{split}
\end{equation}
with $1\leq i\leq j\leq N$. These $\bar\CC$ type multiplets are  originate from tensor product of two $\bar\CB$ type multiplets
\begin{equation}
\bar\CB_{r_1(0,0)}\otimes\bar\CB_{r_2(0,0)}\simeq\bar\CB_{r_1+r_2(0,0)}\oplus\bar\CC_{r_1+r_2-1(\half,0)}\oplus\cdots,
\end{equation}
and
\begin{equation}
\bar\CB_{r_1(0,0)}\otimes\bar\CB_{r_2(\half,0)}\simeq\bar\CB_{r_1+r_2(\half,0)}\oplus\bar\CC_{r_1+r_2-1(0,0)}\oplus\cdots.
\end{equation}
Since these are not chiral ring relations we will not discuss them any further.

\textbf{Example 1}: Consider $\CT^{(A_1, A_2)}[u_1^2]$ which is the $(A_1, A_2)$ theory deformed by $\CO_{\bar\CE_{r_0(0,0)}}=u_1^2$ with $r_0=r(u_1)=\frac{6}{5}$ \cite{Xie:2016hny,Buican:2016hnq}. Some of the protected operator spectrum are
\begin{itemize}
\item Chiral multiplets $u_1$, $\lambda_{1,\alpha}$ and $S_1$,
\item The $\CN=1$ stress tensor multiplet $\hat{\CC}_{(\half,\half)}$,
\item A multiplet of type $\bar\CC_{\frac{1}{6}(\half,0)}$,
\end{itemize}
with the chiral ring relation
\begin{itemize}
\item $u_1^2=0$, $u_1\lambda_{1,\alpha}=0$,
\item $\lambda_{1,\alpha}\lambda_1^\alpha+u_1S_1=0$, $\lambda_{1,\alpha} S_1=0$, and $S_1^2=0$,
\end{itemize}
The existence of $\bar\CC_{\frac{1}{6}(\half,0)}$ multiplet and the first pair of relations was obtained in \cite{Buican:2016hnq} by using the Schur index, while the existence of the second pair of relations was argued in \cite{Xie:2016hny}. Our method reproduce the previous result on $(A_1, A_2)$ and can be easily generalized to any AD theories whose full superconformal indices are known.

\textbf{Example 2}:  Now consider $\CT^{(A_1, A_4)}[u_2]$ with $r_0=r(u_2)=\frac{10}{7}$. The chiral ring of this theory is generated by
\begin{itemize}
\item Chiral multiplets $u_1(\frac{12}{5})$, $\lambda_{1\alpha}(\frac{9}{5})$, $S_1(\frac{6}{5})$, $S_2(\frac{9}{5})$,
\end{itemize}
where the number in the bracket is the IR conformal dimension of the corresponding chiral operator. The index tells us that there is a relation for the operator with smallest dimension $S_1$,
\begin{equation}
S_1^2+u_1=0.
\end{equation}
The $u_1$ term in the relation is necessary because $\Delta(S_1)=\frac{6}{5}$ is smaller than $1.407$ and constraints from bootstrap\cite{Poland:2015mta} tells us that $S_1^2$ can never be $0$. Actually one can see that for any of the IR $\CN=1$ theories obtained by our method, if the conformal dimension of the lowest chiral operator is smaller than $1.407$, there is always other operators in the spectrum with the same conformal dimension twice as the smallest chiral operator, hence our result is consistent with bootstrap bounds.

\subsubsection{Exact marginal deformations} If the UV $\CN=2$ theory has multiple deformations with the same quantum number, the IR theory could have exact marginal deformations. The  marginal operator is in the $\bar\CB_{2(0,0)}$ multiplet and its single letter index has a leading contribution $pq$ to the full index. However, if the IR theory has global symmetries, their conserved currents are in $\hat{\CC}_{(0,0)}$ multiplets which have opposite single letter index comparing to $\bar\CB_{2(0,0)}$ multiplets. Therefore the coefficient of the $pq$ term in the full index is the number of marginal operators minus the number of conserved currents
\begin{equation}
\#~pq-\mathrm{term}=\#~\mathrm{marginal~operators}-\#~\mathrm{conserved~currents}.
\end{equation}
At a generic point on the conformal manifold when all flavor symmetries are broken, the coefficient of the $pq$ term is the actual dimension of  the conformal manifold.

For example, in $(A_1, A_6)$ theory, $u_2^2$ and $u_1u_3$ have the $\CN=2$ $U(1)_r$ charge $\frac{22}{9}$, and in general we can deform the theory with $\CO=au_1u_3+bu^2_2$. After imposing the relation $\CO=0$ we still expect one marginal operator in the IR. Since this theory has no flavor symmetry, this marginal operator can not recombine with any conserved current, hence remains exact marginal. Computation of the full $\CN=1$ index shows that the coefficient of the $pq$ term is exact $1$, hence confirm the fact that the theory has one exact marginal operator in the IR.

In general, $u^2_i$ for $1<i<N$ has the same $U(1)_r$ charge and conformal dimension with $u_{i-1}u_{i+1}$, $u_{i-2}u_{i+2}$, and etc. If one deforms the theory with $\CO_i=u^2_i+a_1u_{i-1}u_{i+1}+a_2u_{i-2}u_{i+2}+\cdots$ with $1<i<N$, the IR theory will have $\mathrm{min}(i-1,N-i)$ exact marginal deformations, where $\mathrm{min}(a,b)$ is the minimum between $a$ and $b$. This leaves a series of $\CN=1$ theories with conformal manifolds. 

One application of such theories is to notice that the IR conformal dimension of $u_j$ with $1\leq j\leq N$ under the deformation $\CO_i=u^2_i+u_{i-1}u_{i+1}+u_{i-2}u_{i+2}+\cdots$ is $\frac{3(N+j+1)}{2(N+i+1)}$ therefore the gap in the spectrum is
\begin{equation}
\Delta_{j+1}-\Delta_{j}=\frac{3}{2(N+i+1)}\sim\frac{3}{2N},
\end{equation}
when $N\gg 1$, hence the theory has a dense spectrum in the sense of \cite{Agarwal:2019crm,Agarwal:2020pol,Agarwal:2020polWIP} at large $N$. Our construction provides first examples of $\CN=1$ theory with both a dense spectrum and a conformal manifold\footnote{We thank Jaewon Song for the discussion.}.

\subsection{$(A_1, A_{2N+1})$ results}
\label{sec:A1A2N+1}

\subsubsection{$\CN=2$ index}

The full index of $(A_1, A_{2N+1})$ AD theories can be computed again using the method  in \cite{Maruyoshi:2016tqk, Maruyoshi:2016aim, Agarwal:2016pjo}. 
First few orders of the single letter indices for $N=1,2,3$ are
\begin{equation}
\CI^{s.l.}_{(A_1, A_3)}=\CI_{\bar\CE_{\frac{4}{3}}}+\CI_{\hat{\CB}_{1}}\chi_1(a)+\CI_{\hat{\CC}_{0(0,0)}}-\CI_{\bar\CB_{1,\frac{4}{3}(0,0)}}\chi_1(a)-\CI_{\bar\CB_{1,\frac{5}{3}(0,0)}}-\CI_{\hat{\CB}_2}-\CI_{\hat{\CB}_3}\chi_1(a)+\cdots,
\end{equation}
\begin{equation}
\begin{split}
\CI^{s.l.}_{(A_1, A_5)}=&\CI_{\bar\CE_{\frac{5}{4}}}+\CI_{\bar\CE_{\frac{3}{2}}}+\CI_{\hat{\CB}_{1}}+\CI_{\hat{\CC}_{0(0,0)}}+\CI_{\hat{\CB}_{\frac{3}{2}}}(a+\frac{1}{a})-\CI_{\hat{\CB}_3}\\
&+\CI_{\bar\CB_{1,\frac{5}{4}(0,0)}}-\CI_{\bar\CB_{1,\frac{3}{2}(0,0)}}-\CI_{\bar\CB_{1,\frac{7}{4}(0,0)}}-\CI_{\bar\CB_{1,2(0,0)}}-(\CI_{\bar\CB{\frac{3}{2},\frac{3}{2}(0,0)}}+\CI_{\bar\CB{\frac{3}{2},\frac{5}{4}(0,0)}})(a+\frac{1}{a})+\cdots,
\end{split}
\end{equation}
and
\begin{equation}
\begin{split}
\CI^{s.l.}_{(A_1, A_7)}=&\CI_{\bar\CE_{\frac{6}{5}}}+\CI_{\bar\CE_{\frac{7}{5}}}+\CI_{\bar\CE_{\frac{8}{5}}}+\CI_{\hat{\CB}_{1}}+\CI_{\hat{\CC}_{0(0,0)}}+\CI_{\hat{\CB}_{2}}(a+\frac{1}{a})-\CI_{\hat\CB_{4}}\\
&-\CI_{\bar\CB_{1,\frac{6}{5}(0,0)}}-\CI_{\bar\CB_{1,\frac{7}{5}(0,0)}}-\CI_{\bar\CB_{1,\frac{8}{5}(0,0)}}-\CI_{\bar\CB_{1,\frac{9}{5}(0,0)}}-\CI_{\bar\CB_{1,\frac{10}{5}(0,0)}}-\CI_{\bar\CB_{1,\frac{11}{5}(0,0)}}\\
&-(\CI_{\bar\CB_{2,\frac{6}{5}(0,0)}}+\CI_{\bar\CB_{2,\frac{7}{5}(0,0)}}+\CI_{\bar\CB_{2,\frac{8}{5}(0,0)}})(a+\frac{1}{a})+\cdots.
\end{split}
\end{equation}
And we conjecture the single letter index of $(A_1, A_{2N+1})$ theory always starts with
\begin{equation}
\label{eq:A1AoddsingleLetterIndex}
\begin{split}
\CI^{s.l.}_{(A_1, A_{2N+1})}=&\sum_{i=1}^N\CI_{\bar\CE_{\frac{N+i+2}{N+2}}}+\CI_{\hat{\CB}_1}+\CI_{\hat{\CC}_{0(0,0)}}+\CI_{\hat{\CB}_{\frac{N+1}{2}}}(a+\frac{1}{a})\\
&-\CI_{\hat\CB_{N+1}}-\sum_{i=0}^{2N-1}\CI_{\bar\CB_{1,\frac{N+i+3}{N+2}(0,0)}}-\left(\sum_{i=1}^N\CI_{\bar\CB_{\frac{N+1}{2},\frac{N+i+2}{N+2}}}\right)(a+\frac{1}{a})+\cdots.
\end{split}
\end{equation}
In the single letter index, the $\hat{\CB}_1$ multiplets contains the Higgs branch operator (moment map of the flavor symmetry). In $(A_1, A_3)$ case the flavor symmetry enhances to $SU(2)$, otherwise the flavor symmetry is $U(1)$.

From the index one immediately sees that at smaller conformal dimension, the spectrum of AD theories consists of $\bar\CE_{r(0,0)}$ type multiplets, the stress tensor multiplet $\hat{\CC}_{0(0,0)}$, $\hat{\CB}_1$ multiplets containing the flavor symmetry current and two more $\hat{\CB}_{\frac{N+1}{2}}$ multiplets charged $\pm1$ under the $U(1)$ flavor symmetry. Primary operators of  $\hat{\CB}_1$ and $\hat{\CB}_{\frac{N+1}{2}}$ multiplets are generators of the Higgs branch chiral ring. On the other hand, there are interesting relations as well. From the $-\CI_{\hat\CB_{N+1}}$ term in the single letter index, there is a relation on Higgs branch
\begin{center}
\begin{tcolorbox}\begin{equation}
\left(\hat\CB^+_{\frac{N+1}{2}}\otimes\hat\CB^+_{\frac{N+1}{2}}\oplus\hat{\CB}_1^{\otimes N+1}\right)_{\hat\CB_{N+1}}=0,
\end{equation}\end{tcolorbox}\end{center}
In terms of generators of the Higgs branch chiral, the relation has the form
\begin{equation}
X^+_{\frac{N+1}{2}}X^-_{\frac{N+1}{2}}+X^{N+1}_1=0.
\end{equation}
Here $X_1$ is the bottom component (primary operator) of the $\hat\CB_1$ multiplet, and $X^{\pm}_{\frac{N+1}{2}}$ are the bottom components of $\hat\CB_{\frac{N+1}{2}}$ with  $U(1)$ flavor charges $\pm1$. Similar to Coulomb branch case, we use $X$'s to denote  primary operators which are generators of the Higgs branch chiral ring to separate them from the primary operator of a generic $\hat\CB_R$ multiplet.

On the other hand, recall the tensor product decomposition
\begin{equation}
\hat\CB_R\otimes\bar\CE_r\simeq\bar\CB_{R,r}+\cdots,
\end{equation}
and the bottom component of $\bar\CB_{R,r}$ is simply the product of bottom components of $\hat\CB_R$ and $\bar\CE_r$.
Therefore $-\CI_{\CB_{1,\frac{N+3}{N+2}(0,0)}}$ leads to the relation $\left(\hat{\CB}_1\otimes\bar\CE_{\frac{N+3}{N+2}}\right)_{\CB_{1,\frac{N+3}{N+2}(0,0)}}=0$, i.e. $X_1u_1=0$. This relations means that the moment map of the $U(1)$ flavor symmetry and the Coulomb branch operator with the lowest dimension can not obtain non-zero vev at the same time. In general, $-\CI_{\bar\CB_{1,r(0,0)}}$ terms in the single letter index implies the relation
\begin{center}
\begin{tcolorbox}
\begin{equation}
\label{eq:A1Aoddrelations1}
\hat{\CB}_1\otimes\bar\CE_{\frac{N+3}{N+2}}=0,
\end{equation}
\end{tcolorbox}\end{center}
\begin{center}
\begin{tcolorbox}
\begin{equation}
\hat{\CB}_1\otimes\bar\CE_{r}+\sum_{r'\in\CA, r+1-r'\in\CA}\left(\bar\CE_{r'}\otimes\bar\CE_{r+1-r'}\right)_{\bar\CB_{1,r(0,0)}}=0,\ \mathrm{for}\,\,r=\frac{N+4}{N+2},\frac{N+5}{N+2},\cdots,\frac{2N+2}{N+2},
\end{equation}
\end{tcolorbox}\end{center}
\begin{center}
\begin{tcolorbox}
\begin{equation}
\sum_{r'\in\CA, r+1-r'\in\CA}\left(\bar\CE_{r'}\otimes\bar\CE_{r+1-r'}\right)_{\bar\CB_{1,r(0,0)}}=0,\ \mathrm{for}\,\,r=\frac{2N+3}{N+2},\frac{2N+4}{N+2},\cdots,\frac{4N+2}{N+2},
\end{equation}
\end{tcolorbox}\end{center}
where $(\cdots)_{\bar\CB_{1,r(0,0)}}$ means picking up the $\CB_{1,r(0,0)}$ in the tensor product decomposition, and $\CA=\{\frac{N+3}{N+2},\frac{N+4}{N+2},\cdots,\frac{2N+2}{N+2}\}$ is the set of all  $U(1)_r$-charges of Coulomb branch operators. Last but not least $-\left(\sum_{i=1}^N\CI_{\bar\CB_{\frac{N+1}{2},\frac{N+i+2}{N+2}}}\right)(a+\frac{1}{a})$ in the single letter index indicates relations between $\hat\CB_{\frac{N+1}{2}}^\pm$ and $\bar\CE_r$ 
\begin{center}
\begin{tcolorbox}\begin{equation}
\label{eq:A1Aoddrelations2}
\left(\hat\CB_{\frac{N+1}{2}}^\pm\otimes\bar\CE_{r}\right)_{\CB_{\frac{N+1}{2},r}}=0,\ \mathrm{for}\,\,r=\frac{N+3}{N+2},\frac{N+4}{N+2},\cdots,\frac{2N+2}{N+2},
\end{equation}\end{tcolorbox}\end{center}
or in terms of generators of the Higgs and Coulomb branch chiral ring
\begin{equation}
X^{\pm}_{\frac{N+1}{2}}u_i=0,\,\mathrm{for}\,\ 1\leq i\leq N,
\end{equation}
which state that operators $X^\pm_{\frac{N+1}{2}}$ in the Higgs branch and Coulomb branch operators $u_i$ can not have non zero vev at the same time.

When $N=1$, the relation simplifies to 
\begin{equation}
\begin{split}
&\hat\CB_1^{\pm,0}\otimes\bar\CE_{\frac{4}{3}}=0,\\
&\left(\bar\CE_{\frac{4}{3}}\otimes\bar{\CE}_{\frac{4}{3}}\right)_{\bar\CB_{1,\frac{5}{3}(0,0)}}=0.
\end{split}
\end{equation}
The first relation tells us that there is no mixed branch in $(A_1, A_3)$ theory. However, This is not the case when $N>1$ and mixed branch is allowed in such cases.

\subsubsection{$\CN=1$ index and chiral ring}
\label{sec:N1deformA1Aodd}

For $(A_1, A_{2N+1})$ theories, again the relevant $\CN=1$ deformation $\CO_{\bar\CE_{r_0(0,0)}}$ is chosen from $\{u_i,u_iu_j\},~i,j=1,\ldots,N$ or any linear combination of such operators with the same quantum number. 

{\bf $\CN=2$ multiplets as $\CN=1$ multiplets:} Beside $\bar\CE_r$, $\hat\CC_{0(0,0)}$ and $\bar\CB_{1,r(0,0)}$ multiplets, there is also $\hat\CB_R$ multiplet in $(A_1, A_{2N+1})$ theories. After the $\CN=1$ deformation, the index of $\hat\CB_R$ multiplet becomes
\begin{equation}
\begin{split}
\CI_{\hat\CB_R}\left((p,q,(pq)^{1-\frac{1}{r_0}}\right)
&=\frac{(pq)^{R\left(1-\frac{1}{r_0}\right)}}{(1-p)(1-q)}
-\frac{(pq)^{R\left(1-\frac{1}{r_0}\right)+\frac{1}{r_0}}}{(1-p)(1-q)}\\
&=\CI_{\bar\CB_{2R\left(1-\frac{1}{r_0}\right)(0,0)}}
+\CI_{\bar\CC_{2(R-1)\left(1-\frac{1}{r_0}\right)(0,0)}}.
\end{split}
\end{equation}
Hence a $\hat\CB_{R}$ multiplet decomposes to an $\CN=1$ chiral multiplet $X$ and an $\CN=1$ $\bar\CC$ type multiplet and other multiplets which do not contribute to our index.
In particular, when $R=1$, we have 
\begin{equation}
\CI_{\hat\CB_1}\left((p,q,(pq)^{1-\frac{1}{r_0}}\right)
=\CI_{\bar\CB_{2\left(1-\frac{1}{r_0}\right)(0,0)}}+\CI_{\hat\CC_{(0,0)}},
\end{equation}
which means that the $\hat\CB_1$ mutliplet has the following decomposition in terms of $\CN=1$ multiplets,
\begin{equation}
\hat\CB_1\simeq\CI_{\hat\CC_{(0,0)}}\oplus\bar\CB_{2\left(1-\frac{1}{r_0}\right)(0,0)}\oplus\CB_{-2\left(1-\frac{1}{r_0}\right)(0,0)},
\end{equation}
which contains an $\CN=1$ chiral, an $\CN=1$ anti-chiral and an $\CN=1$ conserved current multiplet.

{\bf Chiral ring relations: } We first get the $\CN=1$ single letter index of $\CT^{(A_1, A_{2N+1})}[\CO_{\bar\CE_{r_0(0,0)}}]$ theory  by plugging in $t=(pq)^{1-\frac{1}{r_0}}$ in equation \ref{eq:A1AoddsingleLetterIndex}:
\begin{equation}
\begin{split}
&\CI^{s.l.}_{(A_1, A_{2N+1})}\left(p,q,(pq)^{1-\frac{1}{r_0}}\right)\\
=&\sum_{i=1}^N\CI_{\bar\CB_{\frac{2}{r_0}\frac{N+i+2}{N+2}(0,0)}}+\CI_{\bar\CB_{\frac{2}{r_0}\frac{N+i+2}{N+2}+1-\frac{2}{r_0}(\half,0)}}+\CI_{\bar\CB_{\frac{2}{r_0}\frac{N+i+2}{N+2}+2-\frac{4}{r_0}(0,0)}}\\
&+\CI_{\hat{\CC}_{(0,0)}}+\CI_{\hat{\CC}_{(\half,\half)}}+\CI_{\bar\CC_{1-\frac{2}{r_0}(\half,0)}}-\CI_{\bar\CB_{3-\frac{2}{r_0}(\half,0)}}\\
&+\left(\CI_{\bar\CB_{2\left(1-\frac{1}{r_0}\right)(0,0)}}+\CI_{\hat\CC_{(0,0)}}\right)+\left(\CI_{\bar\CB_{(N+1)\left(1-\frac{1}{r_0}\right)(0,0)}}+\CI_{\bar\CC_{(N-1)\left(1-\frac{1}{r_0}\right)(0,0)}}\right)(a+\frac{1}{a})\\
&-\left(\CI_{\bar\CB_{2(N+1)\left(1-\frac{1}{r_0}\right)(0,0)}}-\CI_{\bar\CC_{2N\left(1-\frac{1}{r_0}\right)(0,0)}}\right)\\
&-\sum_{i=0}^{2N-1}\left(\CI_{\bar\CB_{\frac{2}{r_0}\frac{N+i+3}{N+2}+2-\frac{2}{r_0}(0,0)}}+\CI_{\bar\CB_{\frac{2}{r_0}\frac{N+i+3}{N+2}+3-\frac{4}{r_0}(\half,0)}}
+\CI_{\bar\CB_{\frac{2}{r_0}\frac{N+i+3}{N+2}+4-\frac{6}{r_0}(0,0)}}\right)\\
&-\sum_{i=0}^{2N-1}\left(\CI_{\bar\CC_{\frac{2}{r_0}\frac{N+i+3}{N+2}(0,0)}}
+\CI_{\bar\CC_{\frac{2}{r_0}\frac{N+i+3}{N+2}+1-\frac{2}{r_0}(\half,0)}}+\CI_{\bar\CC_{\frac{2}{r_0}\frac{N+i+3}{N+2}+2-\frac{4}{r_0}(0,0)}}\right)+\cdots.
\end{split}
\end{equation}
Similarly to previous case, we can extract the generators of the chiral ring
\begin{center}
\begin{tcolorbox}
\begin{itemize}
\item {\bf Generator set 1:} $u_i$, $\lambda_{i,\alpha}$ and $S_i$ for $1\leq i\leq N$, with $R_{IR}$ charges $\frac{2}{r_0}\frac{N+i+2}{N+2}$, $\frac{2}{r_0}\frac{N+i+2}{N+2}+1-\frac{2}{r_0}$, and $\frac{2}{r_0}\frac{N+i+2}{N+2}+2-\frac{4}{r_0}$,
\item {\bf Generator set 2:} $X$ with $R_{IR}$ charge $2-\frac{2}{r_0}$, and $X^\pm$ with $R_{IR}$ charges $(N+1)(2-\frac{2}{r_0})$, $U(1)$ flavor charge $\pm1$.
\end{itemize}
\end{tcolorbox}\end{center}
The first set of generators coming from $\CN=1$ decomposition of $\bar\CE_r$ multiplets containing generators of the Coulomb branch chiral ring, while the second set of generators from the $\CN=1$ chiral multiplets in the $\hat\CB$ multiplets containing generators of the Higgs branch chiral ring.

One also gets the following chiral ring relations from different $\CN=2$ multiplets and relations: from $\CN=2$ stress tensor multiplet
\begin{center}
\begin{tcolorbox}
\begin{itemize}
\item {\bf Relation 1:} $\CO_{\bar\CE_{r_0(0,0)}}=0$,
\item {\bf Relation 2:} $Q_{1\alpha}\CO_{\bar\CE_{r_0(0,0)}}=0$.
\end{itemize}
\end{tcolorbox}
\end{center}
From $\CN=2$ Higgs branch relations,
\begin{center}
\begin{tcolorbox}
\begin{itemize}
\item {\bf Relation 3:} $X^+X^-+X^{N+1}+\cdots=0$
\end{itemize}
\end{tcolorbox}
\end{center}
where $+\cdots$ means any other operators with the same quantum number. 
From $\CN=2$ relations \ref{eq:A1Aoddrelations1} and \ref{eq:A1Aoddrelations2}
\begin{center}
\begin{tcolorbox}
\begin{itemize}
\item {\bf Relation 4:} $Xu_1=0$, $X\lambda_{1\alpha}=0$, and $XS_1=0$.
\item {\bf Relation 5:} $X^\pm u_i=0$, $X^\pm \lambda_{i\alpha}=0$, $X^\pm S_i=0$, for $1\leq i\leq N$,
\end{itemize}
\end{tcolorbox}
\end{center}
New relation $3$, $4$ and $5$ appear because of the Higgs branch of $(A_1, A_{2N+1})$ theory.
And last but not least from other $\CN=2$ relations
\begin{center}
\begin{tcolorbox}
\begin{itemize}
\item {\bf Relation 6:} $Xu_i+\lambda_{j\alpha}\lambda_k^\alpha+u_jS_k+u_kS_j+\cdots=0$, 
\item {\bf Relation 7:} $X\lambda_{i\alpha}+\lambda_{j\alpha}S_k+\lambda_{k\alpha}S_j+\cdots=0$,
\item {\bf Relation 8:} $XS_i+S_jS_k=0$, 
\end{itemize}
\end{tcolorbox}
\end{center}
for $2\leq i\leq N$, and any pairs of $(j,k)$ with $j+k=i$,
\begin{center}
\begin{tcolorbox}
\begin{itemize}\item {\bf Relation 9:} $\lambda_{j\alpha}\lambda_k^\alpha+u_jS_k+u_kS_j+\cdots=0$, 
\item {\bf Relation 10:} $\lambda_{j\alpha}S_k+\lambda_{k\alpha}S_j+\cdots=0$, 
\item {\bf Relation 11:} $XS_i+S_jS_k=0$, 
\end{itemize}
\end{tcolorbox}
\end{center}
for any pairs of $(j,k)$ with $j+k=N+1,N+2,\cdots,2N$. Relations 6 - 11 which include both chiral multiplets $X$ and $u$, $\lambda_\alpha$ and $S$ are generalization of chiral ring relation 3, 4 and 5 in $(A_1, A_{2N})$ theories. Similar to $(A_1, A_{2N})$ cases, each distinct value of $j+k$ leads to exact one relation and $\cdots$ means one should include all possible operators with the same quantum number.

\subsection{$(A_1, D_{2N+1})$ and $(A_1, D_{2N})$ theories}

We briefly sketch results for $(A_1, D_{2N+1})$ and $(A_1, D_{2N})$ theories as they are similar to $(A_1, A_{2N+1})$ theories. Using the index formula in \cite{Agarwal:2016pjo} one can read off the $\CN=2$ chiral ring generators and relations.

 For $(A_1, D_{2N+1})$ theories,  generators of chiral rings are
\begin{itemize}
\item Three $\hat\CB_1$ multiplets: $\hat\CB_1^0$, $\hat\CB_1^+$, and $\hat\CB_1^-$ which contain moment map $X^0$, $X^{\pm}$ of the flavor $SU(2)$,
\item The stress tensor multiplet $\hat\CC_{0(0,0)}$,
\item $N$ $\bar\CE_{r}$ multiplets: $\bar\CE_{\frac{2N+2i}{2N+1}}$, for $1\leq i\leq N$ which contains $N$ Coulomb branch operators $u_i$.
\end{itemize}
$\CN=2$ relations are
\begin{center}
\begin{tcolorbox}\begin{itemize}
\item Higgs branch chiral ring relation: $\left(\hat\CB_1^+\otimes\hat\CB_1^-\oplus\hat\CB_1^0\right)_{\hat\CB_{2}}=0$, or in terms of moment maps $X^+X^-+(X^0)^2=0$,
\item Mixed branch relation 1: $\left(\hat\CB_{1}^{\pm}\otimes\bar\CE_{\frac{2N+2i}{2N+1}}\right)_{\bar\CB_{1,\frac{2N+2i}{2N+1}(0,0)}}=0$, or in terms of operators $X^{\pm}u_i=0$, for $1\leq i\leq N$,
\item Mixed branch relation 2:  $\left(\hat\CB_{1}^{0}\otimes\bar\CE_{\frac{2N+2}{2N+1}}\right)_{\bar\CB_{1,\frac{2N+2}{2N+1}(0,0)}}=0$, or in terms of operators $X^{0}u_1=0$,
\item Other relations: $\left(\hat\CB_{1}^{0}\otimes\bar\CE_{r_k}\oplus \sum_{r_i,r_j}\bar\CE_{r_i}\otimes\bar\CE_{r_j}\right)_{\bar\CB_{1,r(0,0)}}=0$, for $r=\frac{2N+4}{2N+1},\cdots,\frac{6N-1}{2N+1}$.
\end{itemize}\end{tcolorbox}
\end{center}
One can then read off the corresponding $\CN=1$ chiral ring generators and relations after deforming the theory with $\CO_{\bar\CE_{r_0(0,0)}}$.

For $(A_1, D_{2N})$ theory, the story is similar except that the generators of the Higgs branch chiral ring are enlarged and have four $\hat\CB_1$ multiplets which contain moment maps of $SU(2)\times U(1)$ flavor symmetry, and four $\hat\CB_{\frac{N}{2}}$ multiplets whose bottom components transform as $\mathbf{2}_{1}\oplus \mathbf{2}_{-1}$ under the flavor $SU(2)\times U(1)$.

\subsection{Examples}

We summarize properties of some examples of $\CN=1$ theories deformed from AD theories. The $a$ and $c$ anomalies are shown in table \ref{table:acanomalies}. The IR theory$\CT^{(A_1, A_4)}[u_2]$  has the smallest $a$ and $c$ anomalies in all cases.
\begin{table}
\begin{center}
\begin{tabular}{c|c|c|c|c}
$\CN=2$ theory & deformation $\CO_{\bar\CE_{r_0(0,0)}}$ & $r_0$ & $a(\CT,\CO)$ & $c(\CT,\CO)$ \\\hline
$(A_1, A_2)$ & $u_1^2$ & $\frac{12}{5}$ & $\frac{263}{768}\approx 0.342$ & $\frac{271}{768}\approx 0.353$\\
\hline
\multirow{4}{*}{$(A_1, A_4)$} & $u_2$ & $\frac{10}{7}$ & $\frac{644}{2000}\approx 0.316$ & $\frac{683}{2000}\approx 0.342$\\
  & $u^2_1$ & $\frac{16}{7}$ & $\frac{281}{512}\approx 0.744$ & $\frac{389}{512}\approx 0.760$\\
  & $u_1u_2$ & $\frac{18}{7}$ & $\frac{1013}{1296}\approx 0.782$ & $\frac{1031}{1296}\approx 0.796$\\
  & $u_2^2$ & $\frac{20}{7}$ & $\frac{6369}{8000}\approx 0.796$ & $\frac{6469}{8000}\approx 0.809$\\\hline
  \multirow{6}{*}{$(A_1, A_6)$} & $u_3$ & $\frac{14}{9}$ & $\frac{234}{343}\approx 0.682$ & $\frac{3891}{5488}\approx 0.709$\\
  & $u^2_1$ & $\frac{20}{9}$ & $\frac{37107}{32000}\approx 1.160$ & $\frac{37707}{32000}\approx 1.178$\\
  & $u_1u_2$ & $\frac{22}{9}$ & $\frac{1179}{968}\approx 1.218$ & $\frac{2391}{1936}\approx 1.235$\\
  & $u_1u_3+u_2^2$ & $\frac{8}{3}$ & $\frac{2559}{2048}\approx 1.250$ & $\frac{2591}{2048}\approx 1.265$\\
    & $u_2u_3$ & $\frac{26}{9}$ & $\frac{44379}{35152}\approx 1.262$ & $\frac{22443}{17576}\approx 1.277$\\
      & $u_3^2$ & $\frac{28}{9}$ & $\frac{110871}{87808}\approx 1.263$ & $\frac{112047}{87808}\approx 1.276$\\\hline
        \multirow{7}{*}{$(A_1, A_8)$} & $u_3$ & $\frac{16}{11}$ & $\frac{1563}{2048}\approx 0.763$ & $\frac{1627}{2048}\approx 0.794$\\
  & $u_4$ & $\frac{18}{11}$ & $\frac{695}{648}\approx 1.072$ & $\frac{713}{648}\approx 1.100$\\
  & $u_1^2$ & $\frac{24}{11}$ & $\frac{1213}{768}\approx 1.579$ & $\frac{1229}{768}\approx 1.600$\\
  & $u_1u_2$ & $\frac{26}{11}$ & $\frac{29091}{17576}\approx 1.655$ & $\frac{29429}{17576}\approx 1.674$\\
  & $u_1u_3+u_2^2$ & $\frac{28}{11}$ & $\frac{1335}{784}\approx 1.703$ & $\frac{1349}{784}\approx 1.720$\\
  & $u_1u_4+u_2u_3$ & $\frac{30}{11}$ & $\frac{5189}{3000}\approx 1.730$ & $\frac{5239}{3000}\approx 1.746$\\
    & $u_2u_4+u^2_3$ & $\frac{32}{11}$ & $\frac{28527}{16384}\approx 1.741$ & $\frac{28783}{16384}\approx 1.757$
\end{tabular}
\end{center}
\caption{\label{table:acanomalies}$a$ and $c$ anomalies of AD theory $\CT$ with $\CN=1$ deformation $\CO_{\bar\CE_{r_0(0,0)}}$. $r_0$ is the $U(1)_r$ charge of $\CO_{\bar\CE_{r_0(0,0)}}$. We include only deformations which do not make any chiral multiplets break unitarity bound.}
\end{table}

\begin{table}
\begin{center}
\begin{tabular}{c|c|c|c}
$\CN=2$ theory & \begin{tabular}{c}deformation\\ $\CO_{\bar\CE_{r_0(0,0)}}$ \end{tabular} & \begin{tabular}{c}$\#$ of\\ independent\\ chirals\end{tabular} & independent chirals \\\hline
$(A_1, A_2)$ & $u_1^2$ & 4 & $u_1(1)$, $\lambda_{1\alpha}(\frac{7}{6})$, $S_1(\frac{4}{3})$, $u_1S_1(\frac{7}{3})$ \\
\hline
\multirow{13}{*}{$(A_1, A_4)$} & $u_2$ & 5 & $u_1(\frac{8}{5})$, $\lambda_{1\alpha}(\frac{6}{5})$, $S_1(\frac{4}{5})$, $u_1S_1(\frac{12}{5})$, $S_2(\frac{6}{5})$  \\\cline{2-4}
  & $u^2_1$ & $\infty$  & \begin{tabular}{c}$u_1$, $\lambda_{1\alpha}$, $S_1$, $u_1S_1$\\ $u^n_2$, $u^n_2\lambda_{2\alpha}$, $u^n_2S_2$,\\ $u_1u^n_2$, $u^n_2S_1$, $u_1u^n_2S_1$, $u_1u^n_2S_2$, \\ $u^n_2\lambda_{1\alpha}$, $u_2^n(\lambda_{1\alpha}\lambda_{2\beta}-\frac{1}{2}\epsilon_{\alpha\beta}\lambda_{1\gamma}\lambda_{2}^\gamma)$, $S_1\lambda_{2\alpha}$\end{tabular}  \\\cline{2-4}
  & $u_1u_2$ & $\infty$ & \begin{tabular}{c}$u_1^m$, $u_1^m\lambda_{1\alpha}$, $u_1^mS_1$,\\ $u_2^n$, $u_2^n\lambda_{2\alpha}$, $u^n_2S_2$,\\ $u_1^mS_2$, $u^n_2S_1$, \\$u_1\lambda_{2\alpha}$,  $\lambda_{1\alpha}\lambda_{2\beta}-\frac{1}{2}\epsilon_{\alpha\beta}\lambda_{1\gamma}\lambda_{2}^\gamma$, $S_1\lambda_{2\alpha}$ \end{tabular} \\\cline{2-4}
  & $u_2^2$ & $\infty$ & \begin{tabular}{c}$u_1^m$, $u_1^m\lambda_{1\alpha}$, $u_1^mS_1$,\\ $u_2$, $\lambda_{2\alpha}$, $S_2$, $u_2S_2$,\\ $u_1^mu_2$, $u_1^mS_2$, $u_1^mu_2S_1$, $u^m_1u_2S_2$,\\ $u^m_1\lambda_{2\alpha}$, $u_1^m(\lambda_{1\alpha}\lambda_{2\beta}-\frac{1}{2}\epsilon_{\alpha\beta}\lambda_{1\gamma}\lambda_{2}^\gamma)$, $S_1\lambda_{2\alpha}$\end{tabular}  \\
\end{tabular}
\end{center}
\caption{\label{table:independentchirals}List of independent chirals of AD theory $\CT$ with $\CN=1$ deformation $\CO_{\bar\CE_{r_0(0,0)}}$. The number inside bracket after each operator  is its $R_{IR}$ charge.}
\end{table}

Another interesting example is $\CT^{(A_1, A_6)}[u_2]$ with $r_0=\frac{4}{3}$. Under this deformation, the naive $R_{IR}$ charge of $u_1$ is $\frac{2}{3}$ which is the $R_{IR}$ charge of a free chiral. All the other chiral multiplet has $R_{IR}$ charge above the unitarity bound $\frac{2}{3}$. The $a$ and $c$ anomalies are
\begin{equation}
\begin{split}
a(\CT^{(A_1, A_6)}[u_2])&=\frac{93}{256}=a(\CT^{(A_1, A_2)}[ u^2_1])+\frac{1}{48},\\
c(\CT^{(A_1, A_6)}[u_2])&=\frac{101}{256}=c(\CT^{(A_1, A_2)}[ u^2_1])+\frac{1}{24}.
\end{split}
\end{equation}
This indicates that the IR theory $\CT^{(A_1, A_6)}[u_2]$ is $\CT^{(A_1, A_2)}[ u^2_1]$ plus a free chiral multiplet. This statement can also be checked by matching the $\CN=1$ index. Similarly, the IR theory $\CT^{(A_1, A_{10})}[u_3]$ is $\CT^{(A_1,A_4)}[u_1u_2]$ plus a free chiral.

In some cases, only finitely many chiral multiplets remain in the IR. We summarize some results on independent chiral multiplets in table \ref{table:independentchirals}.

%
%
%
%
%
%
%

\newpage

\section{Free chirals in the IR}
\label{sec:freechiraldeform}

In this section we study a set of particular deformation which leads to free chiral multiplets in the IR. In many cases the Schur index of the $\CN=2$ theory is known and we can obtain the $\CN=1$ index of the deformed theory using the method described in section \ref{sec:Schur}. We see that the IR Schur indices always look like products of $\CN=1$ chiral multiplets. 

\subsection{$(A_1, G)$ theories}

\subsubsection{$(A_1, A_{2N})$: one free chiral}

In $(A_1, A_{2N})$ theories, the Coulomb branch operator $u_1$  has dimension $\frac{2N+4}{2N+3}$. After deforming with $\CO_{\bar\CE_{r_0(0,0)}}=u_1$, the $\CN=1$ full index is computed by substituting $t=(pq)^{1-\frac{1}{r_0}}=(pq)^{\frac{1}{2N+4}}$ into formula \ref{eq:A1A2NfullN2index} and the result is \footnote{If one set $t=(pq)^{\frac{1}{2(N+2)}}$ naively in  \ref{eq:A1A2NfullN2index}, the result will be zero because of $\Gamma(pq)$ term in the index. To get the correct result, we first set $t=(pq)^{\frac{1}{2(N+2)}+\epsilon}$ in \ref{eq:A1A2NfullN2index}, compute the integral using the residue formula and then take the $\epsilon\rightarrow0^+$ limit.}
\begin{equation}\Gamma((pq)^\frac{1}{N+1}),\end{equation}
This is the same index of a free chiral multiplet with $U(1)_R$ charge ${2\over N+1}\leq\frac{2}{3}$. Since the naive $R_{IR}$ charge of the chiral multiplet $S_1$ is ${2\over N+1}$, and we suspect that the IR theory of this particular deformation is just this single free chiral multiplet. 

Let us now look at the Shur index of the $(A_1, A_{2N})$ theory  \cite{Song:2017oew, Xie:2019zlb}, and they take the following simple form:
 \begin{equation}\pe{\frac{q^2-q^{2N+2}}{(1-q)(1-q^{2N+3})}},\end{equation} which equals to the $\CN=1$ index with the substitution $p\rightarrow q^{\frac{1}{r_0-1}}=q^{2N+3}$, which we might call it a reduced $\mathcal{N}=1$ index with just one fugacity $q$. Let's now check that 
 the above index is equal to that of a free chiral with $U(1)_R$ charge ${2\over N+1}$.
 Recall the $\CN=1$ index of a chiral multiplet with $U(1)_R$ charge $r$ is
\begin{equation}
\CI^{\CN=1}_{chiral}(p,q)=\Gamma((pq)^\frac{r}{2})=\pe{\frac{(pq)^\frac{r}{2}-(pq)^{1-\frac{r}{2}}}{(1-p)(1-q)}}.
\end{equation}
After replacing $p\rightarrow q^{\frac{1}{r_0-1}}$, we have
\begin{equation}
\CI^{\CN=1}_{chiral}(q^{\frac{1}{r_0-1}},q)=\pe{\frac{(q^{\frac{r_0}{r_0-1}})^\frac{r}{2}-(q^{\frac{r_0}{r_0-1}})^{1-\frac{r}{2}}}{(1-q^{\frac{1}{r_0-1}})(1-q)}}=\pe{\frac{q^2-q^{2N+2}}{(1-q)(1-q^{2N+3})}}
\end{equation}
Here we use the fact that $r_0={2N+4\over 2N+3},~r={2\over N+2}$. The use of Shur index also indicates that the IR theory is just a free chiral \footnote{See \cite{Bolognesi:2015wta} for the discussion for small $N$.}.

\begin{table}
\begin{center}
\begin{tabular}{c|c|c|c}
$\CN=2$ $u_i$ & $u_i$ & $\lambda_{i,\alpha}$ & $S_i$\\
\hline
$\frac{2N+2+2i}{2N+3}$ & $\frac{2(N+1+i)}{N+2}$ & $\frac{N+2i+1}{N+2}$ & $\frac{2i}{N+2}$
\end{tabular}
\end{center}
\caption{\label{table:A1A2NfreeChiralDim}Conformal dimensions of $\CN=1$ chiral multiplets from the $\CN=2$ $\CE_r$ multiplets. The first column is the dimension of the $\mathcal{N}=2$ Coulomb branch operator, while subsequent columns are dimensions of each $\CN=1$ chirals in the IR limit. In the table, $i$ runs from $1$ to $N$.}
\end{table}

The naive $R_{IR}$-charge of three $\CN=1$ chiral multiplets of $\CT^{(A_1, A_{2N})}[u_1]$ from $\CN=2$ $\bar\CE_{r(0,0)}$ multiplets  are listed in table \ref{table:A1A2NfreeChiralDim}. From the table one sees immediately that chirals $S_i$ with $1\leq i\leq \lfloor\frac{N+2}{3}\rfloor$\footnote{$\lfloor x\rfloor$ means the largest integer which is no greater than $x$.} violate the unitarity bound. One common procedure is to assume that operators which violate the unitarity bound become free, and there is an extra $U(1)$ symmetry which couples only with these operators and the correct IR $U(1)_R$ charge should include this $U(1)$ so these operators have the correct R-charge for a free chiral \cite{Kutasov:2003iy}. The true $a$ and $c$ anomalies should be replacing the naive contributions of chirals violating the unitarity bound with contributions of free chirals,
\begin{equation}
\begin{split}
a^{true}&=a^{trial}-a(M_R)+a^{free}\\
c^{true}&=c^{trial}-c(M_R)+c^{free}
\end{split}
\end{equation}
where $M_R$ is the multiplet violating the unitarity bound, $a(M_R)$ and $c(M_R)$ are the anomaly contribution from $M_R$ with naive R-charge $R$, and $a^{free}=\frac{1}{48}$ and $c^{free}=\frac{1}{24}$ are anomalies of a free chiral.

Direct computation shows that the naive $a$ and $c$ anomalies of $\CT^{(A_1, A_{2N})}[u_1]$ is
\begin{equation}
\begin{split}
a(\CT^{(A_1, A_{2N})}[u_1])&=-\frac{3}{16}\frac{N(N^2-2N-2)}{(N+2)^3},\\
c(\CT^{(A_1, A_{2N})}[u_1])&=-\frac{1}{8}\frac{N(N^2-5N-5)}{(N+2)^3},
\end{split}
\end{equation}
which are the same as the $a$ and $c$ anomalies of the $S_1$ chiral with the naive $R_{IR}$ listed in table \ref{table:A1A2NfreeChiralDim}. This implies that in the IR, $S_1$ becomes free, and  the correct $a$ and $c$ anomalies should be,
\begin{equation}
\begin{split}
a^{true}(\CT^{(A_1, A_{2N})}[u_1])&=-\frac{3}{16}\frac{N(N^2-2N-2)}{(N+2)^3}-a(S_1)+\frac{1}{48}=\frac{1}{48},\\
c^{true}(\CT^{(A_1, A_{2N})}[u_1])&=-\frac{1}{8}\frac{N(N^2-5N-5)}{(N+2)^3}-c(S_1)+\frac{1}{24}=\frac{1}{24},
\end{split}
\end{equation}
which is the anomalies for a theory of one free chiral multiplet. In the IR, we only see a single free chiral multiplet, and one might be curious about the fate of other chiral multiples which simply disappear in the index and $a,c$ anomaly. We will
use the chiral ring relation to explain why these extra chirals vanish.

For example, if we deform $(A_1, A_4)$ theory with $\CO_{\bar\CE_{r_0(0,0)}}=u_1$ and ignore the unitarity bound temporarily, the IR theory has the chiral ring relations
\begin{equation}
\begin{split}
&u_1=0,\,\lambda^a_1=0,\\
&u_1S_1+\lambda_{1\alpha}\lambda^\alpha_1+u_2=0,\\
&\lambda_{1\alpha}S_1+\lambda^a_2=0,\\
&S^2_1+S_2=0. 
\end{split}
\end{equation}
From the chiral ring relation, we see that $u_2, S_2, \lambda_2$ can be expressed in terms of the polynomial of $S_1$ and so is no longer the generators of the chiral ring. One can similarly show that only $S_1$ survives in the chiral ring. 

\subsubsection{$(A_1, D_{2N+1})$: three free chirals}

The smallest Coulomb branch operator $u_1$ of $(A_1, D_{2N+1})$ theory has dimension $\frac{2N+2}{2N+1}$, while it Schur index is 
\begin{equation}
\label{eq:SchurA1D2N+1}
\begin{split}
&\pe{\frac{(q-q^{2N+1})\chi_1(z)}{(1-q)(1-q^{2N+1})}}\\
=&\pe{\frac{z^2q-z^{-2}q^{2N+1}}{(1-q)(1-q^{2N+1})}}
\pe{\frac{q-q^{2N+1}}{(1-q)(1-q^{2N+1})}}
\pe{\frac{z^{-2}q-z^2q^{2N+1}}{(1-q)(1-q^{2N+1})}}
\end{split}
\end{equation} where $\chi_1(z)$ is the character of the adjoint representation of $SU(2)$\cite{Song:2017oew, Xie:2019zlb}. After deforming the theory with $u_1$, the UV Schur index \ref{eq:SchurA1D2N+1} is the same as the $\CN=1$ index of the IR theory $\CT^{(A_1, D_{2N+1})}[u_1]$ with $p\rightarrow q^{\frac{1}{r_0-1}}=q^{2N+1}$, and it is also the same as three chiral multiplets with $U(1)_R$-charge $\frac{1}{N}$ which have an flavor $SU(2)$ symmetry. Therefore it is quite possible that the IR theory is just a theory of three free chirals.

\begin{table}
\begin{center}
\begin{tabular}{c|c|c|c||c}
$\CN=2$ ${\cal E}_i$ & $u_i$ & $\lambda_{i,\alpha}$ & $S_i$ & $X_a$\\
\hline
$\frac{2N+2i}{2N+1}$ & $\frac{2(N+i)}{N+1}$ & $\frac{N+2i}{N+1}$ & $\frac{2i}{N+1}$ & $\frac{1}{N+1}$
\end{tabular}
\end{center}
\caption{\label{table:A1D2N+1freeChiralDim}Conformal dimensions of $\CN=1$ chiral multiplets of $\CT^{(A_1, D_{2N+1})}[u_1]$. The first column is the dimension of the $\mathcal{N}=2$ multiplets corresponding to generators of Coulomb branch chiral ring of $(A_1, D_{2n+1}$ theory, while subsequent columns are dimensions of each $\CN=1$ chirals in the IR limit, and the last coulomb is the R-charge of the chiral from $\hat{\CB}_1$ multiplet in the IR limit. In the table, $i$ runs from $1$ to $N$. }
\end{table}

The conformal dimensions from naive $R_{IR}$-charge of chiral multiplets of $\CT^{(A_1, D_{2N+1})}[u_1]$ is listed in table \ref{table:A1D2N+1freeChiralDim}. One sees that all $X_a$ from $\CN=2$ moment maps and $S_i$'s with $1\leq i\leq\lfloor\frac{N+1}{3}\rfloor$ violate the unitarity bound. The IR $a$ and $c$ anomalies are
\begin{equation}
\begin{split}
a_{\CN=1}&=\frac{9}{32}\frac{N(2N^2-2N-1)}{(N+1)^3},\\
c_{\CN=1}&=\frac{3}{32}\frac{N(4N^2-10N-5)}{(N+1)^3},
\end{split}
\end{equation}
which are exactly three times the $a$ and $c$ anomalies of the $X_a$ multiplet with naive R-charge $\frac{1}{N+1}$. This implies that the three $X_a$'s from moment maps become free chirals in the IR, and the correct $a$ and $c$ anomalies of the IR theory should be
\begin{equation}
\begin{split}
a^{true}_{\CN=1}&=\frac{9}{32}\frac{N(2N^2-2N-1)}{(N+1)^3}-\sum_{a}a(X_a)+\frac{1}{16}=\frac{1}{16},\\
c^{true}_{\CN=1}&=\frac{3}{32}\frac{N(4N^2-10N-5)}{(N+1)^3}-\sum_{a}c(X_a)+\frac{1}{8}=\frac{1}{8},
\end{split}
\end{equation}
therefore the IR theory is a free theory with three chirals. Note that some $S_i$'s also violate unitarity bound when $N$ large enough, however, chiral ring relations tell us $S_i=0$ for any $i$ therefore none of them appear in the IR. For example, in the IR theory $\CT^{(A_1, D_3)}[u_1]$, there is a chiral ring relation
\begin{equation}
X^+X^-+X^2+S_1=0
\end{equation}
among other relations. All $X$'s have $R_{IR}$ charge $\half$ hence violating unitarity bound, and should be removed from the relation and yields $S_1=0$. Deforming the theory with $u_1$ also leads to relation $u_1=\lambda_{1\alpha}=0$, so all the $\mathcal{N}=1$ chiral 
multiplets from $\mathcal{N}=2$ multiplet containing $u_1$ disappear. One might explain the disappearance of other $\mathcal{N}=1$ chiral multiplets from $\CE$-type multiplets in $\CT^{(A_1, D_{2N+1})}[u_1]$ theories using the same reasoning.

\subsubsection{$(A_1, A_{2N+1})$: four free chirals}

In $(A_1, A_{2N+1})$ theory, the conformal dimension of $u_1$ is $\frac{N+3}{N+2}$, and the Schur index can be recast into the product of four terms 
\begin{equation}
\begin{split}
&\pe{\frac{q+q^2+(a+a^{-1})q^{\frac{N+1}{2}}-(a+a^{-1})q^{\frac{N+1}{2}+2}-q^{N+1}-q^{N+2}}{(1-q)(1-q^{N+2})}}\\
=&\pe{\frac{q-q^{N+2}}{(1-q)(1-q^{N+2})}}\pe{\frac{q^2-q^{N+1}}{(1-q)(1-q^{N+2})}}\pe{\frac{aq^{\frac{N+1}{2}}-a^{-1}q^{\frac{N+1}{2}+2}}{(1-q)(1-q^{N+2})}}\pe{\frac{a^{-1}q^{\frac{N+1}{2}}-aq^{\frac{N+1}{2}+2}}{(1-q)(1-q^{N+2})}}.
\end{split}
\end{equation}
When $N=1$ the theory is the same as $(A_1, D_3)$ and the IR theory after deforming with $u_1$ is just the theory three free chiral multiplets because the second term in the product becomes $1$ in this case. 

When $N>1$, the Schur index of $\CT^{(A_1, A_{2N+1})}[u_1]$ (which is the same as the UV Schur index) looks like the index of four chirals with $U(1)_R$-charge $\frac{2}{N+3}$, $\frac{4}{N+3}$, $\frac{N+1}{N+3}$, and $\frac{N+1}{N+3}$. However, the IR theory is more subtle because $\frac{4}{N+3}=\frac{4}{5}>\frac{2}{3}$ for $N=2$, while $\frac{N+1}{N+3}>\frac{2}{3}$ for $N>3$ which means that naive R-charges of these chirals do not violate the unitarity bound so one may question that if these chirals decouple in the IR. Hence we need carefully analyze the $a$ and $c$ anomalies to gain more information on IR.

The naive IR $a$ and $c$ anomalies of $\CT^{(A_1, A_{2N+1})}[u_1]$ are
\begin{equation}
\begin{split}
a_{\CN=1}&=-\frac{3}{8}\frac{N^3-4N^2-6N+3}{(N+3)^3},\\
c_{\CN=1}&=-\frac{1}{8}\frac{2N^3-20N^2-39N-9}{(N+3)^3},
\end{split}
\end{equation}
which are the same as the sum of $a$ and $c$ anomalies of four chirals with naive R-charge, that is
\begin{equation}
\begin{split}
a_{\CN=1}&=a(X(\frac{2}{N+3}))+a(S_1(\frac{4}{N+3}))+2a(X^+(\frac{N+1}{N+3})),\\
c_{\CN=1}&=c(X(\frac{2}{N+3}))+c(S_1(\frac{4}{N+3}))+2c(X^+(\frac{N+1}{N+3})).
\end{split}
\end{equation}
This is strong evidence that all four chirals become free in the IR even though some of them have naive R-charge which is above the unitarity bound, so the actual $a$ and $c$ anomalies of the IR theory should be
\begin{equation}
\begin{split}
a^{true}_{\CN=1}&=a_{\CN=1}-a(X(\frac{2}{N+3}))-a(S_1(\frac{4}{N+3}))-2a(X^+(\frac{N+1}{N+3}))+4\times \frac{1}{48}=\frac{1}{12},\\
c^{true}_{\CN=1}&=c_{\CN=1}-c(X(\frac{2}{N+3}))-c(S_1(\frac{4}{N+3}))-2c(X^+(\frac{N+1}{N+3}))+4\times \frac{1}{24}=\frac{1}{6}.
\end{split}
\end{equation}
Notice that when $N=1$, $S_1(\frac{4}{N+3})$ has IR R-charge $1$, which allows a mass deformation and becomes massive, leaving three free chirals in this case. Although when $N>3$ the $R_{IR}$ charge of $X^\pm$ from $\hat\CB^{\pm}_{\frac{N+1}{2}}$ is $\frac{N+1}{N+3}$ which do not violate unitarity bound, these operators still become free in the IR.

\subsubsection{$(A_1, D_{2N})$: eight free chirals}

For $(A_1, D_{2N})$ theory, the conformal dimension of $u_1$ is $\frac{N+1}{N}$, and the Schur index is
\begin{equation}
\pe{\frac{(1+\chi_1(z))q+(a+a^{-1})\chi_{\half}(z)q^{\frac{N}{2}}-(a+a^{-1})\chi_{\half}(z)q^{\frac{N+2}{2}}-(1+\chi_1(z))q^{N}}{(1-q)(1-q^{N})}}.
\end{equation}
When $N=2$, the flavor symmetry of the $(A_1, D_4)$ theory enhances to $SU(3)$ and by the same argument as in $(A_1, D_{2N+1})$ case the IR theory is a free theory with eight chiral multiplets.

When $N>2$, the Schur index suggests that the IR theory $\CT^{(A_1, D_{2N})}[u_1]$ has four chirals with naive R-charge $\frac{2}{N+1}$ and four chirals with naive $U(1)_R$-charge $\frac{N}{N+1}$. Computation shows that the naive $\CN=1$ $a$ and $c$ anomalies are the same as $a$ and $c$ anomalies from these chirals
\begin{equation}
\begin{split}
a_{\CN=1}&=4a(X(\frac{2}{N+1}))+4a(X(\frac{N}{N+1})),\\
c_{\CN=1}&=4c(X(\frac{2}{N+1}))+4c(X(\frac{N}{N+1})).
\end{split}
\end{equation}
By the same argument as the previous section, one concludes that the IR theory is a theory with eight free chiral multiplets for $N>2$ as well, and the actual central charges are $a=\frac{1}{6}$ and $c=\frac{1}{3}$. These chirals come from 
$\CN=2$ $\hat{B}$ mutliplets which contain generators of the Higgs branch operators of parent $\mathcal{N}=2$ theory, and the disappearance of $S_1$ operator can be explained using the chiral ring relation $S_1+X^2=0$, here $X$ has IR $U(1)_R$ charge ${2\over N+1}$.

\subsection{$(A_{k-1}, G)$ theories}

We briefly summarize results for $(A_{k-1}, G)$ theories, with $\mathrm{gcd}(k,h^\vee)=1$ where $h^\vee$ is the dual Coxeter number of $G$. The Schur index of $(A_{k-1}, G)$ theories with $G=ADE$ deformed by the smallest Coulomb branch operator  is
\begin{equation}
\label{eq:AkGSchur}
\pe{\frac{\sum_i q^{m_i+1}-\sum_iq^{k+m_i}}{(1-q)(1-q^{k+h^\vee})}}
=\pe{\frac{\sum_i q^{m_i+1}-\sum_iq^{k+h-m_i}}{(1-q)(1-q^{k+h^\vee})}},
\end{equation}
where  $\{m_i\}$'s are exponents of $G$. The equality uses the fact that the set $\{m_i\}$ is the same as $\{h^\vee-m_i\}$. Here we take $k>h^\vee$.

The smallest Coulomb branch operator of $(A_{k-1},G)$ theory has the conformal dimension $\frac{k+h^\vee+1}{k+h^\vee}$, therefore the $\CN=1$ index  is the same as the Schur index \ref{eq:AkGSchur} which also looks like the index \textbf{$\mathrm{rank}\,G$} chiral multiplets with the substitution $p=q^{k+h^\vee}$. By the same logic we conjecture that $\CT^{(A_{k-1}, G)}[u_1]$ is a theory with $\mathrm{rank}\,G$ free chiral multiplets. This fact is also confirmed by the central charge computation.

\subsection{$W^{k'}(\fg,f)$ theories}

Now we consider $\CN=2$ SCFTs corresponds to the vertex operator algebra $W^{k'}(\fg,f)$ with $k'=-h^\vee+\frac{h^\vee}{h^\vee+k}$, $h^\vee$ dual Coxeter number of $\fg$ and $f$ labels a nilpotent orbit of $\fg$. For generic $f$, the Lie group $G$ of $\fg$ has a subgroup $SU(2)\times G_F$ with $G_F$ being the flavor group dtermined by $f$. Under this subgroup the adjoint represetation of $G$ decomposes as
\begin{equation}
adj_{G}=\sum_jV_j\otimes R_j,
\end{equation}
where $V_j$ is the spin $j$ representation of $SU(2)$ and $R_j$ is the corresponding representation of $G_F$. 
The Schur index is
\begin{equation}
\pe{\frac{\sum_jq^{1+j}\chi_{R_j}(z)-q^{k+h^\vee}\sum_jq^{-j}\chi_{R_j}(z)}{(1-q)(1-q^{k+h^\vee})}}.
\end{equation}

Again the smallest conformal dimension of Coulomb branch operators is $\frac{k+h^\vee+1}{k+h^\vee}$. Under this deformation the $\CN=1$ index at $p=q^{k+h^\vee}$ limit looks like an index of $\sum_j\dim R_j$ chirals. Hence we conjecture that the IR theory $\CT^{W^{k'}(\fg,f)}[u_1]$ is a theory with  $\sum_j\dim R_j$ free chirals. Again, central charge computation is consistent with the claim.

\section{Other phenomenon}
\label{sec:otherPheno}

\subsection{Interacting theory plus free chirals}
We have seen that the IR theory of deformed $\mathcal{N}=2$ theory could be interacting $\CN=1$ SCFT, or free theory. Here we study the intermediate case where the IR theory consists of interacting theory plus free chirals.

Let us first look at the IR theory $\CT^{(A_1, A_6)}[u_2]$ of $(A_1, A_6)$. According to equation \ref{eq:A1A2NfullN2index}, the $\CN=2$ index of $(A_1, A_6)$ theory is
\begin{equation}
\begin{split}
\CI^{(A_1, A_{6})}_{\CN=2}
=&\left[\prod_{i=1}^3\frac{\Gamma\left(\left(\frac{pq}{t}\right)^{\frac{2(4+i)}{9}}\right)}{\Gamma\left(\left(\frac{pq}{t}\right)^{\frac{2i}{9}}\right)}\right]\Gamma\left(\left(\frac{pq}{t}\right)^{\frac{1}{9}}\right)^3\\
&\times\frac{\kappa^3}{2^33!}\oint[d\bfz]\prod_{\alpha\in\Delta}\frac{\Gamma\left(\bfz^\alpha\left(\frac{pq}{t}\right)^\frac{1}{9}\right)}{\Gamma(\bfz^\alpha)}
\prod_{w\in R}\Gamma\left(\bfz^w\left(\frac{pq}{t}\right)^\frac{4}{9}t^\half\right)\Gamma\left(\bfz^w\left(\frac{pq}{t}\right)^{-\frac{1}{3}}t^\half\right),
\end{split}
\end{equation}
where $\bfz=\{z_1, z_2, z_3\}$ are fugacities of $Sp(6)$. Deforming the theory with $u_2$ means that we take $t= (pq)^{1/4}$ in the $\CN=2$ index. However, taking this limit naively leads to a $\Gamma\left(pq\right)$ factor in the index and by definition $\Gamma\left(pq\right)=0$.

To take the correct $\CN=1$ limit, we use the similar trick as the previous section. First setting $t= (pq)^{\frac{1}{4}+\epsilon}$, then performing the integral and taking $\epsilon\rightarrow0$ limit in the end. After integrating over $z_1$ and $z_2$ fugacities and rescaling $z_3$, we find that
\begin{equation}
\begin{split}
\CI^{\CT^{(A_1, A_6)}[u_2]}_{\CN=1}
&=\frac{\kappa}{2}\Gamma\left((pq)^\frac{1}{3}\right)
\frac{\Gamma\left((pq)^\frac{1}{12}\right)}{\Gamma\left((pq)^\frac{1}{6}\right)}
\oint dz\frac{\Gamma\left((pq)^\frac{1}{12}z^{\pm2}\right)}{\Gamma\left(z^{\pm2}\right)}\Gamma\left((pq)^\frac{5}{24}z^{\pm}\right)\Gamma\left((pq)^\frac{11}{24}z^{\pm}\right)\\
&=\Gamma\left((pq)^\frac{1}{3}\right)\CI^{(A_1, A_2),u^2_1}_{\CN=1},
\end{split}
\end{equation}
which is the $\CN=1$ index of the IR theory $\CT^{(A_1, A_2)}[u_1^2]$   times the index of the theory of a free chiral multiplet. This result confirms the previous observation that the IR theory $\CT^{(A_1, A_6)}[u_2]$  is the IR theory $\CT^{(A_1, A_2)}[u^2_1]$ theory plus a free chiral multiplet using $a$ and $c$ anomalies. The decoupled free chiral comes from $S_1$ whose naive $R_{IR}$ charge is ${2\over 3}$, which is the same as a free chiral.

It is now easy to look for generalizations: we look for  a $\mathcal{N}=1$ preserving deformation using the operator $\CO_{{\cal E}_{r_0(0,0)}}$,  such that the $S_1$ chiral multiplet becomes free, i.e. we need to solve the equation
\begin{equation}
{3(r_0+r_{min}-2)\over r_0}=1
\end{equation}
Here $r_{min}$ is the smallest Coulomb branch operator charge. For example, if we look at $(A_1, A_{2N})$ theory, and we have $r_{min}={2N+4\over 2N+3}$, and we have 
\begin{equation}
r_0={3N+3\over 2N+3}
\end{equation}
There are operators with this $U(1)_r$ charge in the Coulomb branch if $N$ is odd. 

\subsection{Accidental symmetry}

Interesting things happen when we consider $\CT^{(A_1, A_{2N})}[u_2]$ theory with $N\geq4$. The $U(1)_r$ charge $r_0$ of $u_2$ is $\frac{2N+6}{2N+3}$, so we should take $t=(pq)^{\frac{3}{2N+6}}$ in the $\CN=2$ index to get the index of $\CT^{(A_1, A_{2N})}[u_2]$. We found that after integrating out $z_1, z_2,\cdots,z_{N-1}$ and rescaling $z_N$, the $\CN=1$ index is
\begin{equation}
\begin{split}
\CI^{\CT^{(A_1, A_{2N})}[u_2]}_{\CN=1}
=&\frac{\kappa}{2}\Gamma\left((pq)^\frac{2}{N+3}\right)\Gamma\left((pq)^\frac{3}{N+3}\right)\\
&\times
\frac{\Gamma\left((pq)^\frac{1}{2N+6}\right)}{\Gamma\left((pq)^\frac{1}{N+3}\right)}
\oint dz\frac{\Gamma\left((pq)^\frac{1}{2N+6}z^{\pm2}\right)}{\Gamma\left(z^{\pm2}\right)}\Gamma\left((pq)^\frac{2N-1}{4N+12}z^{\pm}\right)\Gamma\left((pq)^\frac{2N+5}{4N+12}z^{\pm}\right)\\
=&\Gamma\left((pq)^\frac{2}{N+3}\right)\Gamma\left((pq)^\frac{3}{N+3}\right)\CI^N_{\CN=1},
\end{split}
\end{equation}
where we define $\CI^N_{\CN=1}$ to be the integral
\begin{equation}
\CI^N_{\CN=1}=\frac{\kappa}{2}\frac{\Gamma\left((pq)^\frac{1}{2N+6}\right)}{\Gamma\left((pq)^\frac{1}{N+3}\right)}\oint dz\frac{\Gamma\left((pq)^\frac{1}{2N+6}z^{\pm2}\right)}{\Gamma\left(z^{\pm2}\right)}\Gamma\left((pq)^\frac{2N-1}{4N+12}z^{\pm}\right)\Gamma\left((pq)^\frac{2N+5}{4N+12}z^{\pm}\right).
\end{equation}

The above index formula suggests the the IR theory contains two free chiral multiplets and an interacting theory. We try to understand above results using the index.
Assuming that the IR theory $\CT^{(A_1, A_{2N})}[u_2]$ contains two free chiral multiplets (which comes from $S_1$ and $S_2$), 
and the interacting theory whose index is $\CI^N_{\CN=1}$, one can study the generators and relations of this interacting piece by computing the single letter index from $\CI^N_{\CN=1}$. For example, when $N=4$, the  index takes the form
\begin{align}
&\CI^{\CT^{(A_1, A_{8})}[u_2]}_{\CN=1}=PE[\CI_{\bar\CB_{\frac{4}{7}(0,0)}}+\CI_{\bar\CB_{\frac{8}{7}(\half,0)}}+\CI_{\bar\CB_{\frac{6}{7}(0,0)}}+\CI_{\hat\CC_{(0,0)}}- \nonumber\\
&~~~~~~~~~~~~~~~~~~~~\CI_{\bar\CB_{\frac{16}{7}(0,0)}}-\CI_{\bar\CB_{\frac{16}{7}(\half,0)}}-\CI_{\bar\CB_{\frac{18}{7}(\half,0)}}-\CI_{\bar\CB_{\frac{20}{7}(0,0)}}+\CI_{\hat\CC_{(\half,\half)}}+\cdots ].
\end{align}
The naive $R_{IR}$ charge for  chiral operators are listed in table \ref{charge}. From the index, we see that  generators are $S_1, \lambda_1^\alpha, S_2$.  Other chiral operators listed in table \ref{charge} are no longer  generators of the chiral ring as we have following chiral ring relation ${\cal B}= f(\lambda_1^\alpha, S_1, S_2)$. 
There is a contribution from the conserved current multiplet $\hat{\CC}_{(0,0)}$ in the single letter index which indicates that there is an accidental $U(1)$  symmetry in the IR. From the index, we also see that the chiral ring relation for $S_1, \lambda_1^\alpha, S_2$ takes the following form
\begin{align}
& \lambda_1^\alpha \lambda_{1\alpha}+S_1S_2^2=0 \nonumber\\
& \lambda_1^\alpha S_1^2=0 \nonumber\\
& \lambda_1^\alpha S_1S_2=0 \nonumber\\
&  \lambda_1^\alpha \lambda_{1\alpha}S_1+S_1^2S_2^2+S_1^5=0
\label{chiralor}
\end{align}
Next, the index for the interacting part takes the following form 
\begin{align}
&\CI^4_{\CN=1}=PE[\CI_{\bar\CB_{\frac{8}{7}(0,0)}}+\CI_{\bar\CB_{\frac{8}{7}(\half,0)}}+\CI_{\bar\CB_{\frac{10}{7}(0,0)}}+\CI_{\hat\CC_{(0,0)}}- \nonumber\\
&~~~~~~~~~~~~~~~~~~~~\CI_{\bar\CB_{\frac{16}{7}(0,0)}}-\CI_{\bar\CB_{\frac{16}{7}(\half,0)}}-\CI_{\bar\CB_{\frac{18}{7}(\half,0)}}-\CI_{\bar\CB_{\frac{20}{7}(0,0)}}+\CI_{\hat\CC_{(\half,\half)}}+\cdots].
\end{align}
Here  generators and their $R_{IR}$ charges are $({\cal O}, \lambda_\alpha, S)=({8\over 7},{8\over 7},{10\over 7})$.  They might be identified with  chiral operators of the original theory as $({\cal O}, \lambda_\alpha, S)=(S_1^2, \lambda_{1}^\alpha, S_1S_2)$, and the 
chiral ring relation takes the following form 
\begin{align}
& \lambda_1^\alpha \lambda_{1\alpha}=0, \nonumber\\
& \lambda_1^\alpha {\cal O}=0, \nonumber\\
& \lambda_1^\alpha  S=0, \nonumber\\
&  S^2=0.
\label{specialchiral}
\end{align}

\begin{table}
\begin{center}
\begin{tabular}{|c|c|} \hline
$\mathcal{N}=2$ chiral $u_i$ & $\mathcal{N}=1$ chiral $({\cal O}_i,\lambda_i^\alpha, S_i)$) \\ \hline
$u_1$ & $(\left.\frac{12}{7},\frac{8}{7},\frac{4}{7}\right)$ \\ \hline
$u_2$ & $(2,\frac{10}{7},\frac{6}{7})$ \\ \hline
$u_3$ & $(\frac{16}{7},\frac{12}{7},\frac{8}{7})$ \\ \hline
$u_4$ & $(\frac{18}{7},2,\frac{10}{7})$ \\ \hline
\end{tabular}
\end{center}
\caption{$U(1)_R$ charge of various chiral operators of theory $\CT^{(A_1, A_{2N})}[u_2]$.}
\label{charge}
\end{table}

We would like to make several remarks for the above series of theories:
\begin{enumerate}
\item The computation of chiral ring can be generalized to arbitrary $N\geq 4$, and we get the same chiral ring generators and relations as  \ref{specialchiral}. This suggests the interacting theory should be the same one for all $N$ although  generators for the chiral ring have 
different naive $R_{IR}$ charge. Fortunately, there is an accidental $U(1)$ symmetry and so the true $U(1)_R$ symmetry would be the mixing of the naive $R_{IR}$ symmetry and this accidental $U(1)$ symmetry. 
\item When $N=4$ or $5$, $S_2$ does not violate the unitarity bound. One might propose that $S_2$ does not become free, but  we find an inconsistency as there would be no combination of chiral operators give the third chiral ring relation of \ref{chiralor} if we just 
take $S_1$ to be free. 
\item  When $N=3$, there are two differences. Before taking the unitarity bound into consideration,  generators for the chiral ring is $(S_1, \lambda_1^\alpha, S_2)=({2\over3},{7\over 6}, 1)$, as the index takes the form
 \begin{align}
&\CI^{\CT^{(A_1, A_{6})}[u_2]}_{\CN=1}=PE[\CI_{\bar\CB_{\frac{2}{3}(0,0)}}+\CI_{\bar\CB_{\frac{7}{6}(\half,0)}}+\CI_{\bar\CB_{1(0,0)}}+\CI_{\hat\CC_{(0,0)}}- \nonumber\\
&~~~~~~~~~~~~~~~~~~~~\CI_{\bar\CB_{\frac{7}{3}(0,0)}}-\CI_{\bar\CB_{\frac{5}{2}(\half,0)}}-\CI_{\bar\CB_{\frac{13}{6}(\half,0)}}-\CI_{\bar\CB_{\frac{8}{3}(0,0)}}+\CI_{\hat\CC_{(\half,\half)}}+\cdots].
\end{align}
Chiral ring relations take the following form
\begin{align}
& \lambda_1^\alpha \lambda_{1\alpha}+S_1^2S_2=0, \nonumber\\
& \lambda_1^\alpha S_1^2=0, \nonumber\\
& \lambda_1^\alpha S_2=0, \nonumber\\
&  S_1S_2^2+S_1^4=0.
\end{align}
These relations are different from those listed in \ref{chiralor}, and especially the third relation has been changed, as properties of the IR theory.
First there is no accidental $U(1)$ symmetry because the chiral operator $S_2^2$ is  marginal  (has $R_{IR}$ charge $2$), which can be combined with the conserved current of $U(1)$ symmetry (one can also see the fact by noticing that there is no $t^6$ term in the full index). Secondly only $S_1$ becomes free, and  the index for the interacting theory takes the form
 \begin{align}
&\CI^{\CT^{(A_1, A_{2N})}[u_2]}_{\CN=1}=PE[\CI_{\bar\CB_{1(0,0)}}+\CI_{\bar\CB_{\frac{7}{6}(\half,0)}}+\CI_{\bar\CB_{\frac{4}{3}(0,0)}}+\CI_{\hat\CC_{(0,0)}}- \nonumber\\
&~~~~~~~~~~~~~~~~~~~~\CI_{\bar\CB_{\frac{7}{3}(0,0)}}-\CI_{\bar\CB_{\frac{5}{2}(\half,0)}}-\CI_{\bar\CB_{\frac{13}{6}(\half,0)}}-\CI_{\bar\CB_{\frac{8}{3}(0,0)}}+\CI_{\hat\CC_{(\half,\half)}}+\cdots],
\end{align}
and now chiral ring generators might be identified as $({\cal O}, \lambda_\alpha, S)=(S_1^2, \lambda_{1}^\alpha, S_2)$, and chiral ring relations take the form
\begin{align}
& \lambda_1^\alpha \lambda_{1\alpha}+{\cal O}S=0, \nonumber\\
& \lambda_1^\alpha {\cal O}=0, \nonumber\\
& \lambda_1^\alpha  S=0, \nonumber\\
&  S^2=0.
\end{align}
\end{enumerate}

\section{Conclusions}
\label{sec:conc}
It was shown in \cite{Xie:2019aft} that one can get 4d $\mathcal{N}=1$ SCFTs by turning on $\mathcal{N}=1$ preserving deformations of 4d $\mathcal{N}=2$ Argyres-Douglas theories.
AD theories are special in that they contain many interesting $\mathcal{N}=1$ chiral operators ${\cal O}_{\CE_{r(0,0)}}$ with scaling dimension less than three, and so one 
can get interesting IR  $\mathcal{N}=1$ SCFTs by turning on tese relevant dimensions. This is in strong contrast with Lagrangian theory or $T_N$ theory \cite{Benini:2009mz}. 

Since the space of 4d AD theories is now amazingly large \cite{Xie:2012hs,Xie:2015rpa} and one can get a large class of $\mathcal{N}=1$ SCFTs with these deformations. The space of these SCFTs seem 
very useful in learning exact properties of $\mathcal{N}=1$ SCFTs, as many properties of AD theories such as their Coulomb branch spectrum, Higgs branch spectrum, 
superconformal index, and etc have been worked out. Results of parent $\mathcal{N}=2$ SCFTs can thus be used to learn properties of IR SCFTs. The IR properties of $\mathcal{N}=1$ preserving RG flow is a lot more subtle though, and so it is  important to get some exact results about them. 
The index of a 4d $\mathcal{N}=1$ SCFT is a very useful tool. In this paper, we use the known result of $\mathcal{N}=2$ index to get the $\mathcal{N}=1$ index, which 
is then used to learn  properties of IR theory, such as the phase structure, the chiral ring,  the accidental symmetry, and etc. 

We would like to make some further comments on  lessons learned about $\mathcal{N}=1$ dynamics. Here we  start with a $\mathcal{N}=2$ SCFT and have a $\mathcal{N}=1$ preserving RG flow with a candidate IR $U(1)_R$ symmetry:
\begin{enumerate}
\item For some RG flows, many operators violate unitarity bound under the naive $R_{IR}$ symmetry. It was pointed out that in these cases \cite{Kutasov:2003iy}, such operators become free. However, we find 
that this is actually quite subtle. It seems that if the operator with worst violation becomes free, many other chiral operators simply drop out from the chiral ring. This happens because if we take a chiral operator free, the chiral ring relation also changes, and under the modified chiral ring relation, some of the operators become trivial in the chiral ring. In this paper, we found many models where the above phenomenon indeed happens.
This suggests that the key object seems to be the chiral ring, as suggested in \cite{Collins:2016icw}.  
\item Another subtle issue is the accidental symmetry, as we have shown in some models where there is no apparent appearance of extra symmetry. The models studied in this paper shows that one must be very careful about the accidental symmetry of any studies of $\mathcal{N}=1$ theories  using the RG flow.
This issue is often overlooked in the literature and might lead to serious mistakes.
\item The IR phase structure of our RG flow is  quite rich: it could be interacting SCFTs, free theories, interacting theories plus free chirals, and etc. The IR SCFTs could be isolated or have conformal manifold. We have not found the appearance of any gapped 
phase though.
\end{enumerate}

The space of $\mathcal{N}=1$ SCFTs found in \cite{Xie:2019aft} is very large and quite interesting, and these new $\mathcal{N}=1$ theories definitely deserve further studies.

\section*{Acknowledgements}
 DX and Wenbin Yan are supported by Yau Mathematical Science Center at Tsinghua University, the
Young overseas high-level talents introduction plan, national key research and development
program of China (NO. 2020YFA0713000), and NNSF of China with Grant NO: 11847301
and 12047502.

\newpage
\appendix

\section{$\CN=2$ BPS multiplets and their indices}
\label{app:sec:BPSmultiplets}

A generic long multiplet ${\cal A}_{R,r(j_1,j_2)}^{\Delta}$ of the $\NN=2$
superconformal algebra is generated by the action of the eight Poincar\'e supercharges
$\QQ$ and $\tilde{\QQ}$ on a superconformal primary, which by definition is
 annihilated by all  conformal supercharges ${\cal S}$. If  some combination of
the  $\QQ$s  also annihilates the primary, the corresponding multiplet
is shorter and the conformal dimensions of all its members are protected against quantum corrections.
The shortening 
conditions for the $\NN=2$ superconformal algebra were studied in
 \cite{Dobrev:1985qv,Dobrev:1985qz,Dolan:2002zh}.
 We follow the nomenclature of \cite{Dolan:2002zh}, except our $U(1)_r$ charge $r$ is $-r$ of \cite{Dolan:2002zh}. The classification scheme under our notation is summarized in  table \ref{shortening}.
Let us take a moment to explain the notation.
The state $|R,r\rangle^{h.w.}_{(j_1,j_2)}$ is the highest weight state with
$SU(2)_{R}$ spin $R >0$, $U(1)_{r}$ charge $r$,
which can have either sign, and Lorentz quantum numbers  $(j_1,j_2)$.
The multiplet  built on this state is  denoted as $\mathcal{X}_{R,r(j_1,j_2)}$,
where the letter $\mathcal{X}$ characterizes the shortening condition.
The left column of table  \ref{shortening} labels
the condition. 
A superscript on the label  corresponds to the index $\II =1,2$ of the
supercharge that kills the primary:
for example ${\cal B}_{\suup}$ refers
to ${\cal Q}_{\suup\alpha}$. Similarly a ``bar'' on the label refers to the conjugate condition: for example
$\bar{\BB}_{\sudown}$ corresponds to $\tilde Q_{\sudown \, \dot \alpha}$ annihilating the state;
this would result in the short anti-chiral multiplet $\bar{\BB}_{R,r(j_1,0)}$, obeying $\Delta = 2 R +r$.
Note that conjugation reverses the signs of $r$, $j_1$ and $j_2$ in the expression of the conformal dimension. The highest weight state of $\bar\CE_{r(j_1,0)}$ multiplet is annihilated by both $\tilde{Q}_{1\dot{\alpha}}$ and $\tilde{Q}_{2\dot{\alpha}}$, and we call it the $\CN=2$ chiral multiplet with BPS condition $\Delta=r$ in our convention of $U(1)_r$\footnote{In \cite{Dolan:2002zh}, the authors call $\CE_{r(0,j_1)}$ chiral multiplet. Here we call $\bar\CE_{r(j_1,0)}$ chiral to be consistent with the definition of chiral multiplet in the usual $\CN=1$ convention. }.

\begin{table}
\begin{centering}
\begin{tabular}{|c|l|l|l|l|}
\hline 
\multicolumn{4}{|c|}{Shortening Conditions} & Multiplet\tabularnewline
\hline
\hline 
$\BB_{\suup}$  & $\QQ_{\suup\alpha}|R,r\rangle^{h.w.}=0$  & $j_1=0$ & $\Delta=2R-r$  & $\BB_{R,r(0,j_2)}$\tabularnewline
\hline 
$\bar{\BB}_{\sudown}$  & $\tilde{\QQ}_{\sudown \dot{\alpha}}|R,r\rangle^{h.w.}=0$  & $j_2=0$ & $\Delta=2R+r$  & $\bar{\BB}_{R,r(j_1,0)}$\tabularnewline

\hline 
$\EE$  & $\BB_{\suup}\cap\BB_{\sudown}$  & $R=0$  & $\Delta=-r$  & $\EE_{r(0,j_2)}$\tabularnewline
\hline 
$\bar \EE$  & $\bar \BB_{\suup}\cap \bar \BB_{\sudown}$  & $R=0$  & $\Delta=r$  & $\bar \EE_{r(j_1,0)}$\tabularnewline
\hline 
$\hat{\BB}$  & $\BB_{\suup}\cap\bar{B}_{\sudown}$  & $r=0$, $j_1,j_2=0$  & $\Delta=2R$  & $\hat{\BB}_{R}$\tabularnewline
\hline
\hline 
$\CC_{\suup}$  & $\e^{\alpha\beta}\QQ_{\suup\beta}|R,r\rangle_{\alpha}^{h.w.}=0$  &  & $\Delta=2+2j_1+2R-r$  & $\CC_{R,r(j_1,j_2)}$\tabularnewline
 & $(\QQ_{\suup})^{2}|R,r\rangle^{h.w.}=0$ for $j_1=0$  &  & $\Delta=2+2R-r$  & $\CC_{R,r(0,j_2)}$\tabularnewline
\hline 
$\bar \CC_{\sudown}$  & $\e^{\dot\alpha\dot\beta}\tilde\QQ_{\sudown\dot\beta}|R,r\rangle_{\dot\alpha}^{h.w.}=0$  &  & $\Delta=2+2 j_2+2R+r$  & $\bar\CC_{R,r(j_1,j_2)}$\tabularnewline
 & $(\tilde\QQ_{\sudown})^{2}|R,r\rangle^{h.w.}=0$ for $j_2=0$  &  & $\Delta=2+2R+r$  & $\bar\CC_{R,r(j_1,0)}$\tabularnewline
\hline 
  & $\CC_{\suup}\cap\CC_{\sudown}$  & $R=0$  & $\Delta=2+2j_1-r$  & $\CC_{0,r(j_1,j_2)}$\tabularnewline
\hline 
  & $\bar\CC_{\suup}\cap\bar\CC_{\sudown}$  & $R=0$  & $\Delta=2+2 j_2+r$  & $\bar\CC_{0,r(j_1,j_2)}$\tabularnewline
\hline 
$\hat{\CC}$  & $\CC_{\suup}\cap\bar{\CC}_{\sudown}$  & $r=j_1-j_2$  & $\Delta=2+2R+j_1+j_2$  & $\hat{\CC}_{R(j_1,j_2)}$\tabularnewline
\hline   & $\CC_{\suup}\cap\CC_{\sudown}\cap\bar{\CC}_{\suup}\cap\bar{\CC}_{\sudown}$  & $R=0, r=j_1-j_2$ & $\Delta=2+j_1+j_2$  & $\hat{\CC}_{0(j_1,j_2)}$\tabularnewline
\hline
\hline 
$\DD$  & $\BB_{\suup}\cap\bar{\CC_{\sudown}}$  & $-r=j_2+1$  & $\Delta=1+2R+j_2$  & $\DD_{R(0,j_2)}$\tabularnewline
\hline 
$\bar\DD$  & $\bar\BB_{\sudown}\cap{\CC_{\suup}}$  & $r=j_1+1$  & $\Delta=1+2R+j_1$  & $\bar\DD_{R(j_1,0)}$\tabularnewline
\hline 
 & $\EE\cap\bar{\CC_{\sudown}}$  & $-r=j_2+1,R=0$  & $\Delta=-r=1+j_2$  & $\DD_{0(0,j_2)}$\tabularnewline
\hline
  & $\bar\EE\cap{\CC_{\suup}}$  & $r=j_1+1,R=0$  & $\Delta=r=1+j_1$  & $\bar\DD_{0(j_1,0)}$\tabularnewline
\hline
\end{tabular}
\par\end{centering}
\caption{\label{shortening}Shortening conditions
and short multiplets for the  $\NN=2$ superconformal algebra.}
\end{table}

The superconformal index counts with signs the protected states of the theory, up to equivalence 
relations that set to zero all sequences of short multiplets that may in principle recombine 
into long multiplets. 
The recombination rules for ${\cal N}=2$ superconformal algebra are \cite{Dolan:2002zh}
\begin{eqnarray}
{\cal A}_{R,r(j_1,j_2)}^{2R+r+2j_1+2} & \simeq & \CC_{R,r(j_1,j_2)}\oplus\CC_{R+\frac{1}{2},r-\frac{1}{2}(j_1-\frac{1}{2},j_2)}
\label{recomb2}\,,\\
{\cal A}_{R,r(j_1,j_2)}^{2R-r+2 j_2+2} & \simeq & \bar\CC_{R,r(j_1,j_2)}\oplus\bar\CC_{R+\frac{1}{2},r+\frac{1}{2}(j_1,j_2-\frac{1}{2})}
\label{eq:3rd recomb}\,,\\
{\cal A}_{R,j_2-j_1(j_1,j_2)}^{2R+j_1+j_2+2} & \simeq & \hat{\CC}_{R(j_1,j_2)}\oplus\hat{\CC}_{R+\frac{1}{2}(j_1-\frac{1}{2},j_2)}
\oplus\hat{\CC}_{R+\frac{1}{2}(j_1,j_2-\frac{1}{2})}\oplus\hat{\CC}_{R+1(j_1-\frac{1}{2},j_2-\frac{1}{2})}\,.
\label{recomb1}
\end{eqnarray}
The ${\cal C}$, $\bar {\cal C}$ and $\hat {\cal C}$ multiplets
 obey certain ``semi-shortening'' conditions,  while ${\cal A}$ multiplets are generic long multiplets. 
 A long multiplet whose conformal dimension is exactly at the unitarity threshold can be decomposed  into shorter multiplets according to (\ref{recomb2},\ref{eq:3rd recomb},\ref{recomb1}).
  We can formally regard any  multiplet obeying some shortening condition (with the exception of the $\EE$ ($\bar \EE$) types, and $\bar{{\cal D}}_{0(j_{1},0)}$ (${\cal D}_{0(0,j_{2})}$) types)
 as a multiplet of  type ${\cal C}$, $\bar {\cal C}$ or $\hat \CC$  by allowing the spins $j_1$ and $j_2$, whose natural range is over the non-negative
 half-integers,  to  take the value $-1/2$ as well.   The translation is as follows:
\be
\CC_{R,r(-\frac{1}{2},j_2)}\simeq\BB_{R+\frac{1}{2},r-\frac{1}{2}(0,j_2)},\quad \bar\CC_{R,r(j_1,-\frac{1}{2})}\simeq\bar\BB_{R+\frac{1}{2},r+\frac{1}{2}(j_1,0)}\,,
\label{translation}
\ee
\be
\hat{\CC}_{R(-\frac{1}{2},j_2)}\simeq\DD_{R+\frac{1}{2}(0,j_2)},\qquad\quad\qquad\hat{\CC}_{R(j_1,-\frac{1}{2})}\simeq\bar{\DD}_{R+\frac{1}{2}(j_1,0)}\,, 
\ee
\be
\hat{\CC}_{R(-\frac{1}{2},-\frac{1}{2})}\simeq \DD_{R+\frac{1}{2}(0,-\frac{1}{2})} \simeq
\bar{\DD}_{R+\frac{1}{2}(-\frac{1}{2},0)} \simeq \hat{\BB}_{R+1}\,.
\ee
Note how these rules flip statistics: a multiplet with bosonic primary ($j_1+ j_2$ integer) is turned
into a multiplet with fermionic primary ($j_1 +  j_2$ half-odd), and vice versa.  
With these conventions, the rules (\ref{recomb2}, \ref{eq:3rd recomb}, \ref{recomb1}) are the most general recombination rules. 
The  $\EE$ and $\bar \EE$ multiplets never recombine.

The index of the $\CC$ and $\EE$ type multiplets vanishes identically (the choice of supercharge with respect
to which the index is computed, ${\cal Q} =\widetilde  {\cal Q}_{1 \dot -}$,
breaks the symmetry between $\CC$ ($\EE$) and $\bar \CC$ ($\bar \EE$) multiplets).
  The index of all remaining short multiplets can be specified by listing the index of $\bar\CC, \hat \CC$ , $\bar\EE$, ${\cal D}_{0(0,j_{2})}$, and $\bar {\cal D}_{0(j_{1},0)}$ multiplets,
\be
{\cal I}_{\bar\CC_{R,r(j_1, j_2)}}&=&
-(-1)^{2(j_1+ j_2)}p^{j_2+r}q^{j_2+r}t^{R-r-1}\frac{(t-pq)(t-p)(t-q)}{(1-p)(1-q)}\chi_{j_1}\(\sqrt{\frac{p}{q}}\)\,,\nonumber\\
{\cal I}_{{\hat \CC}_{R(j_1, j_2)}}&=&
(-1)^{2(j_1+ j_2)}p^{j_1}q^{j_1}t^{R-j_1+j_2-1}\frac{t-p q}{(1-p)(1-q)}
\left(p^\half q^\half t\chi_{j_1+\half}\left(\sqrt{\frac{p}{q}}\right)
-pq\chi_{j_1}\left(\sqrt{\frac{p}{q}}\right)\right)\,,\nonumber\\
{\cal I}_{\bar\EE_{r(j_1,0)}}&=& (-1)^{2j_1}p^{r-1}q^{r-1}t^{-r} \frac{(t-p)(t-q)}{(1-p)(1-q)}\chi_{j_1}\(\sqrt{\frac{p}{q}}\)\,,\label{indexE}\,\nonumber\\
{\cal I}_{\bar{{\cal D}}_{0(j_{1},0)}} & = & \frac{(-1)^{2j_{1}}(pq/t)^{j_{1}+1}}{(1-p)(1-q)}\times\nonumber\\
&&\qquad\qquad\left((1+t)\chi_{j_{1}}\left(\sqrt{\frac{p}{q}}\right)-\frac{t}{\sqrt{pq}}\chi_{j_{1}+\half}\left(\sqrt{\frac{p}{q}}\right)-\sqrt{pq}\chi_{j_{1}-\half}\left(\sqrt{\frac{p}{q}}\right)\right)\,,
\nonumber\\
{\cal I}_{{\cal D}_{0(0,j_{2})}} & = & \frac{(-1)^{2j_{2}+1}t^{j_{2}}(t-pq)}{(1-p)(1-q)}\,.
\ee
where the Schur polynomial $\chi_{j}\left(\sqrt{\frac{p}{q}}\right)$ gives the character of the spin $j$ representation of $SU(2)$.

The Schur limit is simply $t=q$, and the Schur index of the above short multiplets are
\be
{\cal I}_{\bar\CC_{R,r(j_1, j_2)}}&=&
0\,,\nonumber\\
{\cal I}_{{\hat \CC}_{R(j_1, j_2)}}&=&
(-1)^{2(j_1+j_2)}\frac{q^{2+R+j_1+j_2}}{1-q}\,,\nonumber\\
{\cal I}_{\bar\EE_{r(j_1,0)}}&=& 0\,,\label{indexE}\,\nonumber\\
{\cal I}_{\bar{{\cal D}}_{0(j_{1},0)}} & = & (-1)^{2j_1+1}\frac{q^{j_1+1}}{1-q}\,,\quad
{\cal I}_{{\cal D}_{0(0,j_{2})}}  =  (-1)^{2j_{2}+1}\frac{q^{j_{2}+1}}{1-q}\,.
\ee

\section{$\CN=1$ BPS multiplets and their indices}
\label{app:sec:N1BPS}

In this appendix we summarize some basic facts about $\NN=1$
superconformal representation theory. 
 A generic long multiplet
$\CA^{\Delta}_{r(j_1, j_2)}$ is generated by the action of $4$
Poincar\'e supercharges $\QQ_\alpha$ and ${\widetilde \QQ}_{\dot \alpha}$
on superconformal primary which is by definition is annihilated by
all conformal supercharges $\cal S$.
In table \ref{N1-shortening} we have
summarized  possible shortening and semishortening conditions. Following the usual convention, we call $\bar\CB_{r(j_1,0)}$ chiral multiplets as their highest weight states are annihilated by $\tilde{Q}_{\dot{\alpha}}$'s.

\begin{center}
\begin{table}[!h]
{\small
\begin{centering}
\begin{tabular}{|l|l|l|l|l|}
\hline
\multicolumn{4}{|c|}{Shortening Conditions} & Multiplet\tabularnewline
\hline
\hline
$\BB$ & $\QQ_{\alpha}|r\rangle^{h.w.}=0$ & $j_1=0$ & $\Delta=-\frac{3}{2}r$ & $\BB_{r(0,j_2)}$\tabularnewline
\hline
$\bar{\BB}$ & $\bar{\QQ}_{\dot{\alpha}}|r\rangle^{h.w.}=0$ & $j_2=0$ & $\Delta=\frac{3}{2}r$ & $\bar{\BB}_{r(j_1,0)}$\tabularnewline
\hline
$\hat{\BB}$ & $\BB\cap\bar{\BB}$ & $j_1,j_2,r=0$ & $\Delta=0$ & $\hat{\BB}$\tabularnewline
\hline
\hline
$\CC$ & $\e^{\alpha\beta}\QQ_{\beta}|r\rangle_{\alpha}^{h.w.}=0$ &  & $\Delta=2+2j_1-\frac{3}{2}r$ & $\CC_{r(j_1,j_2)}$\tabularnewline
 & $(\QQ)^{2}|r\rangle^{h.w.}=0$ for $j_1=0$ &  & $\Delta=2-\frac{3}{2}r$ & $\CC_{r(0,j_2)}$\tabularnewline
\hline
$\bar{\CC}$ & $\e^{\dot{\alpha}\dot{\beta}}\bar{\QQ}_{\dot{\beta}}|r\rangle_{\dot{\alpha}}^{h.w.}=0$ &  & $\Delta=2+2j_2+\frac{3}{2}r$ & $\bar{\CC}_{r(j_1,j_2)}$\tabularnewline
 & $(\bar{\QQ})^{2}|r\rangle^{h.w.}=0$ for $j_2=0$ &  & $\Delta=2+\frac{3}{2}r$ & $\bar{\CC}_{r(j_1,0)}$\tabularnewline
\hline
$\hat{\CC}$ & $\CC\cap\bar{\CC}$ & $\frac{3}{2}r=(j_1-j_2)$ & $\Delta=2+j_1+j_2$ & $\hat{\CC}_{(j_1,j_2)}$\tabularnewline
\hline
$\DD$ & $\BB\cap\bar{\CC}$ & $j_1=0,-\frac{3}{2}r=j_2+1$ & $\Delta=-\frac{3}{2}r=1+j_2$ & $\DD_{(0,j_2)}$\tabularnewline
\hline
$\bar{\DD}$ & $\bar{\BB}\cap\CC$ & $j_2=0,\frac{3}{2}r=j_1+1$ & $\Delta=\frac{3}{2}r=1+j_1$ & $\bar{\DD}_{(j_1,0)}$\tabularnewline
\hline
\end{tabular}
\par\end{centering}
} \caption{\label{N1-shortening}Possible shortening conditions 
for the $\NN=1$ superconformal algebra.}
\end{table}
\par\end{center}

At the unitarity threshold, a long multiplet
can decompose into  (semi)short multiplets. 
The splitting rules  are:
\begin{eqnarray*}
\label{app:eq:N1recombine}
\CA_{r(j_1,j_2)}^{2+2j_1-\frac{3}{2}r} & \simeq & \CC_{r(j_1,j_2)}\oplus\CC_{r-1(j_1-\frac{1}{2},j_2)}\\
\CA_{r(j_1,j_2)}^{2+2j_2+\frac{3}{2}r} & \simeq & \bar{\CC}_{r(j_1,j_2)}\oplus\bar{\CC}_{r+1(j_1,j_2-\frac{1}{2})}\\
\CA_{\frac{2}{3}(j_1-j_2)(j_1,j_2)}^{2+j_1+j_2} & \simeq & \hat{\CC}_{(j_1,j_2)}\oplus\CC_{\frac{2}{3}(j_1-j_2)-1,(j_1-\frac{1}{2},j_2)}\oplus\bar{\CC}_{\frac{2}{3}(j_1-j_2)+1,(j_1,j_2-\frac{1}{2})}\end{eqnarray*}
We are using a notation where the
 $\BB$ and $\bar{\BB}$
type multiplets are formally identified with  special cases
of $\CC$ and $\bar{\CC}$ multiplets, as follows
\begin{equation}
\CC_{r(-\frac{1}{2},j_2)}\simeq\BB_{r-1(0,j_2)}\qquad\bar{\CC}_{r(j_1,-\frac{1}{2})}\simeq\bar{\BB}_{r+1(j_1,0)}\,.
\end{equation}
We define the ${\tt L}$eft (${\tt R}$ight) equivalence
class of the multiplet $\CC_{r(j_1,j_2)}(\bar{\CC}_{r(j_1,j_2)})$
as the class of multiplets with the same ${\tt L}$eft (${\tt R}$ight) index. From the splitting
rules, we see that the classes can be labeled as
 $[-r+2j_1,j_2]_{(-)^{2j_1}}^{{\tt L}}$ $([r+2j_2,j_1]_{(-)^{2j_2}}^{{\tt R}})$.
Moreover, ${\cal I}_{[-r+2j_1,j_2]_{-}^{{\tt L}}}^{{\tt L}}=-{\cal I}_{[-r+2j_1,j_2]_{+}^{{\tt L}}}^{{\tt L}}$
and ${\cal I}_{[r+2j_2,j_1]_{-}^{{\tt R}}}^{{\tt R}}=-{\cal I}_{[r+2j_2,j_1]_{+}^{{\tt R}}}^{{\tt R}}$.
The expressions for the indices of the  equivalent classes are
\begin{eqnarray*}
{\cal I}_{[\tilde{r},j_2]_{\pm}^{{\tt L}}}^{{\tt L}} & = & \pm(-)^{2j_2+1}\frac{(pq)^{\half(\tilde{r}+2)}\chi_{j_2}\left(\sqrt{\frac{p}{q}}\right)}{(1-p)(1-q)}\\
{\cal I}_{[\bar{\tilde{r}},j_1]_{\pm}^{{\tt R}}}^{{\tt R}} & = & \pm(-)^{2j_1+1}\frac{(pq)^{\half(\bar{\tilde{r}}+2)}\chi_{j_1}\left(\sqrt{\frac{p}{q}}\right)}{(1-p)(1-q)}\, \\
\II^{R} [\tilde{r},j_2]_{\pm}^{{\tt L}} &  =&  0 \\
 \II^{{\tt L}}[\bar{\tilde{r}},j_1]_{\pm}^{{\tt R}} & = & 0\,.
\end{eqnarray*}

The situation is slightly more involved for the $\hat{\CC}$ and $\DD$ type
multiplets. Unlike the $\BB,\CC$ type multiplets, they contribute both
to $\II^{{\tt L}}$ as well as $\II^{{\tt R}}$. The indices  \cite{Dolan:2008qi} for the different
types of multiplets are collected in table \ref{tab:Indices}.

\newpage

\begin{table}[htbp]
\begin{centering}
\begin{tabular}{|l|l|l|}
\hline
Multiplet & $\II^{{\tt L}}$ & $\II^{{\tt R}}$\tabularnewline
\hline
\hline
$\AA_{r(j_1,j_2)}^{\Delta}$ & $0$ & $0$\tabularnewline
\hline
$\CC_{r(j_1,j_2)}$ & $\II_{[-r+2j_1,j_2]_{(-)^{2j_1}}^{{\tt L}}}^{{\tt L}}$ & $0$\tabularnewline
\hline
$\bar{\CC}_{r(j_1,j_2)}$ & $0$ & $\II_{[r+2j_2,j_1]_{(-)^{2j_2}}^{{\tt R}}}^{{\tt R}}$\tabularnewline
\hline
\hline
$\hat{\CC}_{(j_1,j_2)}$ & $\II_{[\frac{2}{3}j_2+\frac{4}{3}j_1,j_2]_{(-)^{2j_1}}^{{\tt L}}}^{{\tt L}}$ & $\II_{[\frac{2}{3}j_1+\frac{4}{3}j_2,j_1]_{(-)^{2j_2}}^{{\tt R}}}^{{\tt R}}$\tabularnewline
\hline
$\DD_{(0,j_2)}$ & $\II_{[\frac{2}{3}j_2-\frac{4}{3},j_2]_{-}^{{\tt L}}}^{{\tt L}}+\II_{[\frac{2}{3}j_2-\frac{1}{3},j_2-\frac{1}{2}]_{-}^{{\tt L}}}^{{\tt L}}$ & $\II_{[\frac{4}{3}j_2-\frac{2}{3},0]_{+}^{{\tt R}}}^{{\tt R}}$\tabularnewline
\hline
$\bar{\DD}_{(j_1,0)}$ & $\II_{[\frac{4}{3}j_1-\frac{2}{3},0]_{+}^{{\tt L}}}^{{\tt L}}$ & $\II_{[\frac{2}{3}j_1-\frac{4}{3},j_1]_{-}^{{\tt R}}}^{{\tt R}}+\II_{[\frac{2}{3}j_1-\frac{1}{3},j_1-\frac{1}{2}]_{-}^{{\tt R}}}^{{\tt R}}$\tabularnewline
\hline
\end{tabular}
\par\end{centering}

\caption{\label{tab:Indices}Indices $\II^{{\tt L}}$ and $\II^{{\tt R}}$ of the various
short and semi-short multiplets.}

\end{table}

\section{Elliptic Gamma function}
\label{app:sec:eGamma}

The elliptic Gamma function is a two parameter generalization of the Gamma function,
\begin{equation}
\Gamma(z;p,q)\equiv\prod_{j,k\geq0}\frac{1-z^{-1}p^{j+1}q^{k+1}}{1-zp^jq^k}.
\end{equation}
The reader can consult \cite{Diejen:2013,Spiridonov:2008RuMaS,Spiridonov:2005math} for reviews of the elliptic Gamma function and of elliptic hypergeometric mathematics. We will also use the standard condensed notations
\begin{equation}
\begin{split}
&\Gamma(z_1,\cdots,z_k;p,q)\equiv\prod_{j=1}^k\Gamma(z_j;p,q),\\
&\Gamma(z^\pm;p,q)\equiv\Gamma(z;p,q)\Gamma(1/z;p,q),
\end{split}
\end{equation}
and sometimes just omit parameters $p$ and $q$ so $\Gamma(z)\equiv\Gamma(z;p,q)$ without other specifications. There are two useful identities of elliptic Gamma functions
\begin{equation}
\begin{split}
&\Gamma(z^2;p,q)=\Gamma(\pm z,\pm\sqrt{q}z,\pm\sqrt{p}z,\pm\sqrt{pq}z;p,q),\\
&\Gamma(z;p,q)\Gamma(pq/z;p,q)=1.
\end{split}
\end{equation}
Moreover, the residue of $\Gamma(z/a;p,q)$ at $z=a$ is
\begin{equation}
\lim_{a\rightarrow a}(1-z/a)\Gamma(z/a;p,q)=\frac{1}{(p;p)(q;q)},
\end{equation}
with the q-Pochhammer symbol $(z;q)$ defined as
\begin{equation}
(z;q)\equiv\prod_{k=0}^\infty\left(1-zq^k\right).
\end{equation}


\bibliographystyle{JHEP}

\bibliography{ADhigher}

\end{document}